\def\final{}
\def\print{}
\def\squeeze{}

\PassOptionsToPackage{prologue,usenames,dvipsnames,rgb}{xcolor}
\documentclass[10pt,letterpaper,journal,natbib=false]{IEEEtran}
\usepackage[utf8]{inputenc}
\usepackage[T1]{fontenc}
\usepackage[ngerman,english]{babel}
\useshorthands{"}
\addto\extrasenglish{\languageshorthands{ngerman}}
\usepackage{microtype}
\usepackage{amsmath}
\usepackage{amssymb}
\usepackage{amsthm}
\usepackage{graphicx}
\graphicspath{{figures/}}
\usepackage{xspace}
\usepackage{booktabs}
\usepackage{multirow}
\usepackage[ruled,linesnumbered]{algorithm2e}
\usepackage{nicefrac}
\usepackage{pgf}
\usepackage{pgfplots}
\usepackage{pgfplotstable}
\usepackage{tikz}
\usetikzlibrary{%
  arrows,%
  automata,%
  calc,%
  shapes.arrows,%
  shapes.geometric, %
  shapes.misc,%
  chains,%
  graphs,
  matrix,%
  positioning,%
  scopes,%
  decorations.pathmorphing,%
  decorations.markings,
  backgrounds,fit,
  shadows,
  patterns
}
\usetikzlibrary{external}
\usepackage[caption=false]{subfig}
\usepackage[autostyle]{csquotes}
\usepackage{acronym}
\usepackage{ctable}
\usepackage{lipsum}
\usepackage{listings}
\usepackage[defaultlines=2]{nowidow}
\usepackage[shortcuts]{extdash}
\usepackage[inline]{enumitem}
\usepackage{todonotes}
\usepackage{mdframed}

\usepackage{color}
\usepackage[rgb]{xcolor}
\definecolor{ACMBlue}{cmyk}{1,0.1,0,0.1}
\definecolor{ACMYellow}{cmyk}{0,0.16,1,0}
\definecolor{ACMOrange}{cmyk}{0,0.42,1,0.01}
\definecolor{ACMRed}{cmyk}{0,0.90,0.86,0}
\definecolor{ACMLightBlue}{cmyk}{0.49,0.01,0,0}
\definecolor{ACMGreen}{cmyk}{0.20,0,1,0.19}
\definecolor{ACMPurple}{cmyk}{0.55,1,0,0.15}
\definecolor{ACMDarkBlue}{cmyk}{1,0.58,0,0.21}
\usepackage{calc}
\usepackage{newfloat}
    \DeclareFloatingEnvironment[name=\lstlistingname]{listing}
\usepackage{hyperref}
\usepackage{cleveref}
\usepackage{apptools}
    \crefname{subappendix}{\IfAppendix{section}{appendix}}{\IfAppendix{sections}{appendices}s}
    \let\cref\Cref %
\usepackage{scalefnt}
\usepackage{lstautogobble}
\usepackage{etoolbox}

\pdfpageattr {/Group << /S /Transparency /I true /CS /DeviceRGB>>}

\usepackage[style=ieee,backend=biber,giveninits=true,isbn=false,doi=false,minbibnames=1,maxbibnames=2,mincitenames=1,maxcitenames=1]{biblatex}
\AtEveryBibitem{
  \clearfield{venue}
  \clearfield{address}
  \clearname{editor}
}
\DeclareFieldFormat[inproceedings]{pagetotal}{\mkpagetotal[pagination]{#1}}
\renewbibmacro*{chapter+pages}{%
  \printfield{note}%
  \setunit{\bibpagespunct}%
  \printfield{pages}%
  \setunit{\bibpagespunct}%
  \printfield{pagetotal}%
  \newunit}

\presetkeys{todonotes}{inline}{}

\makeatletter
\renewcommand{\todo}[2][color=Cerulean!20]{%
  \ifx\final\empty\else\ifx\print\empty\else%
    \ifx\tikzexternaldisable\undefined\else%
      \tikzexternaldisable%
    \fi
    \@todo[size=\scriptsize,caption={#2},#1]{#2}%
    \ifx\tikzexternalenable\undefined\else%
      \tikzexternalenable%
    \fi
  \fi\fi}
\makeatother

\hypersetup{%
  \ifx\print\empty%
    hidelinks%
  \else%
    colorlinks%
  \fi
}

\def\CC{C\nolinebreak[4]\hspace{-.05em}\raisebox{.4ex}{\relsize{-3}{\textbf{++}}}\xspace}
\def\hipacc{Hipacc\xspace}
\def\anydsl{AnyDSL\xspace}
\def\anyhls{AnyHLS\xspace}
\def\impala{Impala\xspace}

\newcommand{\intextstyle}{\ttfamily\small}
\newcommand{\inlinestyle}{\ttfamily\small}

\expandafter\patchcmd\csname \string\lstinline\endcsname{%
    \leavevmode
    \bgroup
}{%
    \leavevmode
    \ifmmode\hbox\fi
    \bgroup
}{}{%
    \typeout{Patching of \string\lstinline\space failed!}%
}

    {\noindent\ignorespaces\begin{lstlisting}[#1,float,floatplacement=H]}{\end{lstlisting}\noindent\ignorespacesafterend}

\definecolor[named]{whitesmoke}   {rgb}{0.96,0.96,0.96}
\definecolor[named]{codegreen}    {named}{ACMGreen}
\definecolor[named]{codered}      {named}{ACMRed}
\definecolor[named]{codelightblue}{named}{ACMLightBlue}
\definecolor[named]{codedarkblue} {named}{ACMDarkBlue}

\def\lst{\lstinline}
\def\print#1{\pgfmathparse{#1}\pgfmathresult}

\makeatletter
\newenvironment{btHighlight}[1][]
{\begingroup\tikzset{bt@Highlight@par/.style={#1}}\begin{lrbox}{\@tempboxa}}
{\end{lrbox}\bt@HL@box[bt@Highlight@par]{\@tempboxa}\endgroup}

\newcommand\btHL[1][]{%
    \begin{btHighlight}[#1]\bgroup\aftergroup\bt@HL@endenv%
}
\def\bt@HL@endenv{%
    \end{btHighlight}%
    \egroup
}
\newcommand{\bt@HL@box}[2][]{%
  \ifdefined\tikzexternaldisable%
    \tikzexternaldisable%
  \fi%
    \tikz[#1]{%
        \pgfpathrectangle{\pgfpoint{1pt}{0pt}}{\pgfpoint{\wd #2}{\ht #2}}%
        \pgfusepath{use as bounding box}%
        \node[anchor=base west, fill=codelightblue,outer sep=0pt,inner xsep=1pt, inner ysep=0pt, rounded corners=3pt, minimum height=\ht\strutbox+1pt,#1]{\raisebox{1pt}{\strut}\strut\usebox{#2}};
    }%
  \ifdefined\tikzexternalenable%
    \tikzexternalenable%
  \fi%
}

\lstdefinestyle{node}{
    backgroundcolor=,
    language=,
    basicstyle=\tiny\ttfamily,
    morekeywords = {br,neg,or,and,all,any},
    numbers=none,
    mathescape=true,
    frame=none,
    literate={<-}{{$\leftarrow$}}1
}

\lstdefinelanguage{alg}{
    morecomment = [s]{/*}{*/},
    morecomment = [l]{//},
    sensitive = true,
    morekeywords = {for,next,to,step}
}

\lstdefinelanguage{impala}{
    morecomment = [s]{/*}{*/},
    morecomment = [l]{//},
    sensitive = true,
    morekeywords = {i8,i16,i32,i64,u8,u16,u32,u64,f16,f32,f64,bool,int,float,double,extern,struct,as,match,true,false,type,with,let,mut,static,while,in,exit,return,break,continue,if,else,for,do,fn,any,all,extract,shuffle,ballot,enum,stream},
    morekeywords = {@,@@},
    morestring = [b]",
    moredelim=**[is][\btHL]{§}{§},
}

\lstdefinelanguage{metaocaml}{
    sensitive = true,
    morekeywords = {let,in,rec,if,then,else,fun},
    moredelim=**[is][\btHL]{§}{§},
}

\lstdefinelanguage{pseudoml}{
    sensitive = true,
    morekeywords={fun,where,whererec,lambda,let,letrec,in,and,bool,float,int,br,noret},
    literate=%
        {==}{{=}}1
        {!=}{{$\neq$}}1
        {<=}{{$\leq$}}1
        {>=}{{$\geq$}}1
        {->}{{$\rightarrow$}}1
        {<-}{{$\leftarrow$}}1
        {bot}{{$\bot$}}1
        {LAMBDA}{{$\lambda$}}1
}

\lstdefinelanguage{scala}{
    morecomment = [s]{/*}{*/},
    morecomment = [l]{//},
    sensitive = true,
    morekeywords = {val,var,new,with,import,trait,this,def,if,else,Int},
    moredelim=**[is][\btHL]{§}{§},
}

\lstdefinelanguage{scheme}{
    sensitive = true,
    morekeywords={define,filter}
}

\lstdefinelanguage{sierra}{
    morecomment = [s]{/*}{*/},
    morecomment = [l]{//},
    morestring = [b]",
    sensitive = true,
    morekeywords = {uniform,varying,simd,scalar,for_each_active,for_each_unique,current_mask},
    morekeywords = {kernel,uint,mask,skip,true,false,uint32_t,uint64_t,nullptr,return,public,protected,private,template,auto,class,virtual,struct,union,void,this,size_t,volatile,if,else,do,while,case,goto,switch,for,while,bool,typedef,static,const,float,int,short,char,double,break,continue},
    keywords = {[2]define},
    keywordstyle={[2]\color{uds-purple}\bfseries},
    moredelim=**[is][\btHL]{§}{§},
}

\lstdefinelanguage{ssa}{
    sensitive = true,
    morekeywords={fn,bool,float,int,phi,goto,br,return},
    literate=
        {:=}{{$\gets$}}1
        {==}{{=}}1
        {!=}{{$\neq$}}1
        {<=}{{$\leq$}}1
        {>=}{{$\geq$}}1
        {->}{{$\rightarrow$}}1
        {<-}{{$\leftarrow$}}1
        {PHI}{{$\phi$}}1
}

\lstdefinelanguage{terra}{
    morecomment = [s]{/*}{*/},
    morecomment = [l]{//},
    sensitive = true,
    morekeywords = {int,function,if,then,return,else,elseif,terra,end,local},
    moredelim=**[is][\btHL]{§}{§},
}

\lstnewenvironment{lstinlinelisting}[1][]{%
  \noindent%
  \phantom{M}\minipage{\linewidth-.75em}%
  \lstset{frame=none,numbers=none,aboveskip=.5\bigskipamount,belowskip=.5\bigskipamount,backgroundcolor=,#1}%
}{%
  \endminipage%
}
\lstnewenvironment{lstintextlisting}[1][]{%
  \noindent%
  \phantom{M}\minipage{\linewidth-.75em}%
  \lstset{frame=none,numbers=none,basicstyle=\intextstyle,aboveskip=.5\bigskipamount,belowskip=.5\bigskipamount,backgroundcolor=,#1}%
}{%
  \endminipage%
}

\newacro{FPGA}{Field Programmable Gate Array}
\newacro{HLS}{High-Level Synthesis}
\newacro{MUX}{multiplexer}
\newacro{FAC}[F\&C]{Fetch and Calc}
\newacro{CAP}[C\&P]{Calc and Pack}

\newacro{ALU}{Arithmetic Logic Unit}
\newacro{API}{Application Programming Interface}
\newacro{AST}{Abstract Syntax Tree}
\newacro{CGRA}{Coarse-Grained Reconfigurable Architecture}
\newacro{ECC}{Error-Correcting Code}
\newacro{FMA}{Fused Multiply-Add}
\newacro{HDR}{High Dynamic Range}
\newacro{NPP}{NVIDIA Performance Primitives}
\newacro{MAD}{Multiply-Add}
\newacro{SFU}{Special Function Unit}
\newacro{SPL}{Software Product Line}
\newacro{FPS}{Frames Per Second}
\newacro{DSL}{Domain-Specific Language}
\newacro{LUT}{Look-Up Table}
\newacro{BRAM}{Block RAM}
\newacro{DSP}{Digital Signal Processing}
\newacro{PCIe}{PCI Express}
\newacro{MGT}{Multi-Gigabit Transceiver}
\newacro{PLL}{Phase-Locked Loop}
\newacro{SRIO}{Serial RapidIO}
\newacro{SoC}{System-on-a-Chip}
\newacro{MPSoC}{Multi Processor System-on-a-Chip}
\newacro{LoC}{Lines of Code}
\newacro{IP}{Intellectual Property}
\newacro{ISA}{Instruction Set Architecture}
\newacro{PPnR}{Post Place and Route}
\newacro{FF}{Flipflop}
\newacro{DDR3-SDRAM}{Double Data Rate Type Three Synchronous Dynamic Random Access Memory}
\newacro{RAM}{Random Access Memory}
\newacro{SDRAM}{Synchronous Dynamic Random Access Memory}
\newacro{ASIC}{Application-Specific Integrated Circuit}
\newacro{APSoC}{All Programmable System-on-a-Chip}
\newacro{HDL}{Hardware Description Language}
\newacro{SISO}{Single Input Single Output}
\newacro{MISO}{Multiple Input Single Output}
\newacro{MIMO}{Multiple Input Multiple Output}
\newacro{II}{Initiation Interval}
\newacro{FIFO}{First In, First Out}
\newacro{ILP}{Instruction-Level Parallelism}
\newacro{TLP}{Task-Level Parallelism}
\newacro{LLP}{Loop-Level Parallelism}
\newacro{ILLP}{Inner Loop-Level Parallelism}
\newacro{IO}{Input/Output}
\newacro{OLLP}{Outer Loop-Level Parallelism}
\newacro{QoR}{Quality-of-Results}
\newacro{LSGP}{Locally Sequential Globally Parallel}
\newacro{LPGS}{Locally Parallel Globally Sequential}
\newacro{DLP}{Data-Level Parallelism}
\newacro{GPL}{General-Purpose Language}
\newacro{VHDL}{Very high speed Hardware Description Language}
\newacro{RTL}{Register-Transfer Level}
\newacro{VLSI}{Very Large Scale Integration}
\newacro{SSI}{Small Scale Integration}
\newacro{MSI}{Medium Scale Integration}
\newacro{LSI}{Large Scale Integration}
\newacro{CAD}{Computer Aided Design}
\newacro{EDA}{Electronic Design Automation}
\newacro{ESL}{Electronic Systel-Level}
\newacro{CPU}{Central Processing Unit}
\newacro{QoR}{Quality of Results}
\newacro{SLD}{System-Level Design}
\newacro{HSCD}{Hardware/Software Co-Design}
\newacro{DA}{Design Automation}
\newacro{TTL}{Transistor-Transistor Level}
\newacro{DSE}{Design Space Exploration}
\newacro{HLL}{High-Level Programming Language}
\newacro{gcc}{GNU Compiler Collection}
\newacro{AI}{Artificial Intelligence}
\newacro{IDE}{Integrated Development Environment}
\newacro{ASP}{Application-Specific Processor}
\newacro{DAG}{Directed Acyclic Graph}
\newacro{DFG}{Data Flow Graph}
\newacro{CFG}{Control Flow Graph}
\newacro{CDFG}{Control Data Flow Graph}
\newacro{FSMD}{Finite State Machine with Data Path}
\newacro{FSM}{Finite State Machine}
\newacro{LMS}{Lightweight Modular Staging}
\newacro{TCPA}{Tightly-coupled Processor Array}
\newacro{OpenCV}{Open Source Computer Vision}
\newacro{IC}{Integrated Circuit}
\newacro{HPC}{High Performance Computing}
\newacro{SIMD}{Single Instruction Multiple Data}
\newacro{SIMT}{Single Instruction Multiple Threads}
\newacro{MRA}{Multi Resolution Analysis}
\newacro{OOC}{Out-Of-Context}
\newacro{RGBA}{Red Green Blue Alpha}
\newacro{OEM}{Original Equipment Manufacturer}
\newacro{DPRAM}{Dual-Port-RAM}
\newacro{SDK}{Software Development Kit}
\newacro{SPMD}{Single Program Multiple Data}
\newacro{AOC}{\altera Offline Compiler}
\newacro{half-ALM}{half-Adaptive Logic Module}
\newacro{AOCL}{Intel FPGA SDK for OpenCL}
\newacro{LAT}{Latency}
\newacro{SU}{Speedup}
\newacro{ThSU}{Theoretical Speedup}
\newacro{F}{Frequency}
\newacro{TP}{Throughput}
\newacro{ALUT}{Adaptive Look-Up Table}
\newacro{LU}{Logic Utilization}
\newacro{RHS}{Right Hand Side}
\newacro{ROI}{Region Of Interest}
\newacro{SSA}{Static Single Assignment}
\newacro{DFS}{Depth-First Search}
\newacro{TBB}{Threading Building Blocks}
\newacro{IR}{Intermediate Representation}
\newacro{ECU}{Electronic Control Unit}

\acused{DSP}
\acused{LUT}
\acused{BRAM}

\input{figures/src/anyhls_lib.tikz}

\ifx\squeeze\empty

    \AtEveryBibitem{\clearfield{pages}}

    \newcommand{\secbskip}{}
    \newcommand{\secskip}{}

\else
    \newcommand{\secbskip}{}
    \newcommand{\secskip}{}
\fi

\newcommand{\ie}{\mbox{i.e.,}\xspace}
\newcommand{\eg}{\mbox{e.g.,}\xspace}
\newcommand{\floor}[1]{\lfloor{#1}\rfloor}
\newcommand{\ceil}[1]{\lceil{#1}\rceil}

\newcommand{\akif}[1]{\todo[color=yellow,size=\footnotesize]{#1}}

\newcommand{\CopyrightNotice}[2]{%
  \begin{picture}(0,0)(0,0)
    \put(#1){\parbox{\paperwidth-3em}{\sf \center {\scriptsize%
\textcopyright 2020 IEEE.
Personal use of this material is permitted.
Permission from IEEE must be obtained for all other uses, in any current or future media, including reprinting/republishing this material for advertising or promotional purposes,creating new collective works, for resale or redistribution to servers or lists, or reuse of any copyrighted component of this work in other works.
      }}}%
  \end{picture}
  \vspace{#2}
}
\begin{document}


\title{AnyHLS: High-Level Synthesis with Partial Evaluation}

\def\hscd{\IEEEauthorrefmark{3}}
\def\dfki{\IEEEauthorrefmark{2}}
\def\saar{\IEEEauthorrefmark{1}}
\author{\IEEEauthorblockN{
        M.~Akif \"Ozkan\hscd,
        Arsène Pérard-Gayot\dfki,
        Richard Membarth\dfki\saar,
        Philipp Slusallek\dfki\saar,
        Roland Leißa\saar, \\
        Sebastian Hack\saar,
        J\"urgen Teich\hscd, and
        Frank Hannig\hscd}

\IEEEauthorblockA{%
  \hscd%
  Friedrich-Alexander University Erlangen-N\"urnberg (FAU), Germany
}

\IEEEauthorblockA{%
  \saar%
  Saarland University (UdS), Germany
}
\IEEEauthorblockA{%
  \dfki%
  German Research Center for Artificial Intelligence (DFKI), Germany
}
}

\maketitle
\CopyrightNotice{-40,190}{-10pt}

\begin{abstract}
FPGAs excel in low power and high throughput computations, but they are challenging to program.
Traditionally, developers rely on hardware description languages like Verilog or VHDL to specify the hardware behavior at the register-transfer level.
High-Level Synthesis (HLS) raises the level of abstraction, but still requires FPGA design knowledge.
Programmers usually write pragma-annotated C/C++ programs to define the hardware architecture of an application.
However, each hardware vendor extends its own C dialect using its own vendor-specific set of pragmas.
This prevents portability across different vendors.
Furthermore, pragmas are not first-class citizens in the language.
This makes it hard to use them in a modular way or design proper abstractions.

In this paper, we present \anyhls, an approach to synthesize FPGA designs in a modular and abstract way.
\anyhls is able to raise the abstraction level of existing HLS tools by resorting to programming language features such as types and higher-order functions as follows:
It relies on partial evaluation to specialize and to optimize the user application based on a library of abstractions.
Then, vendor-specific HLS code is generated for Intel and Xilinx FPGAs.
Portability is obtained by avoiding any vendor-specific pragmas at the source code.
In order to validate achievable gains in productivity, a library for the domain of image processing is introduced as a case study, and its synthesis results are compared with several state-of-the-art \ac{DSL} approaches for this domain.
\end{abstract}

\IEEEpeerreviewmaketitle

\newsavebox{\mylistingop}
\newsavebox{\mylistinghw}
\begin{lrbox}{\mylistingop}
\begin{lstlisting}[frame=none,basicstyle=\scriptsize\ttfamily,backgroundcolor=]
let sobel_x = @|img, x, y|
    -1 * img.read(x-1, y-1) + 1 * img.read(x+1, y-1) +
    -2 * img.read(x-1, y  ) + 2 * img.read(x+1, y  ) +
    -1 * img.read(x-1, y+1) + 2 * img.read(x+1, y+1);
\end{lstlisting}%
\end{lrbox}

\begin{lrbox}{\mylistinghw}
\begin{lstlisting}[frame=none,basicstyle=\scriptsize\ttfamily,backgroundcolor=]
let input = make_img_mem1d("sandiego.jpg");
let output = make_img_mem1d("output.jpg");
let operator = make_local_op(sobel_x);
with generate(vhls) { operator(input, output) }
\end{lstlisting}%
\end{lrbox}

\begin{figure*}
\begin{mdframed}[backgroundcolor=whitesmoke]
\vspace{-1em}
\noindent
\begin{tabular}{@{}ccc@{}}
    \parbox[c]{0.18\linewidth}{\includegraphics[height=0.128\textwidth,trim={5cm 0 0cm 0},clip]{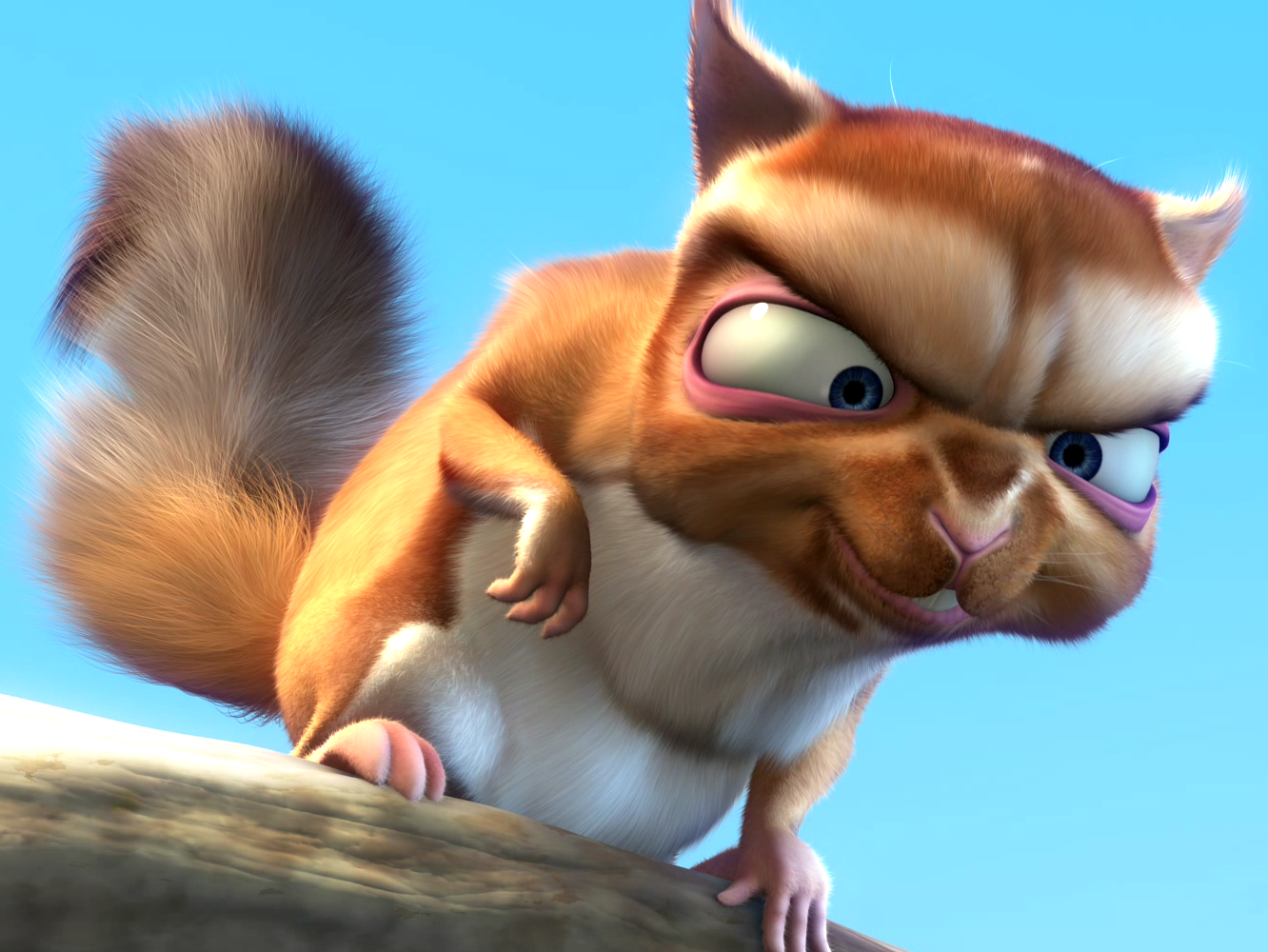}}&
    \parbox[c]{0.58\linewidth}{\vspace{1em}\scalebox{0.65}{\usebox\anyhlsLocalOpDetail}} &
    \parbox[c]{0.14\linewidth}{\includegraphics[height=0.128\textwidth,trim={5cm 0 0cm 0},clip]{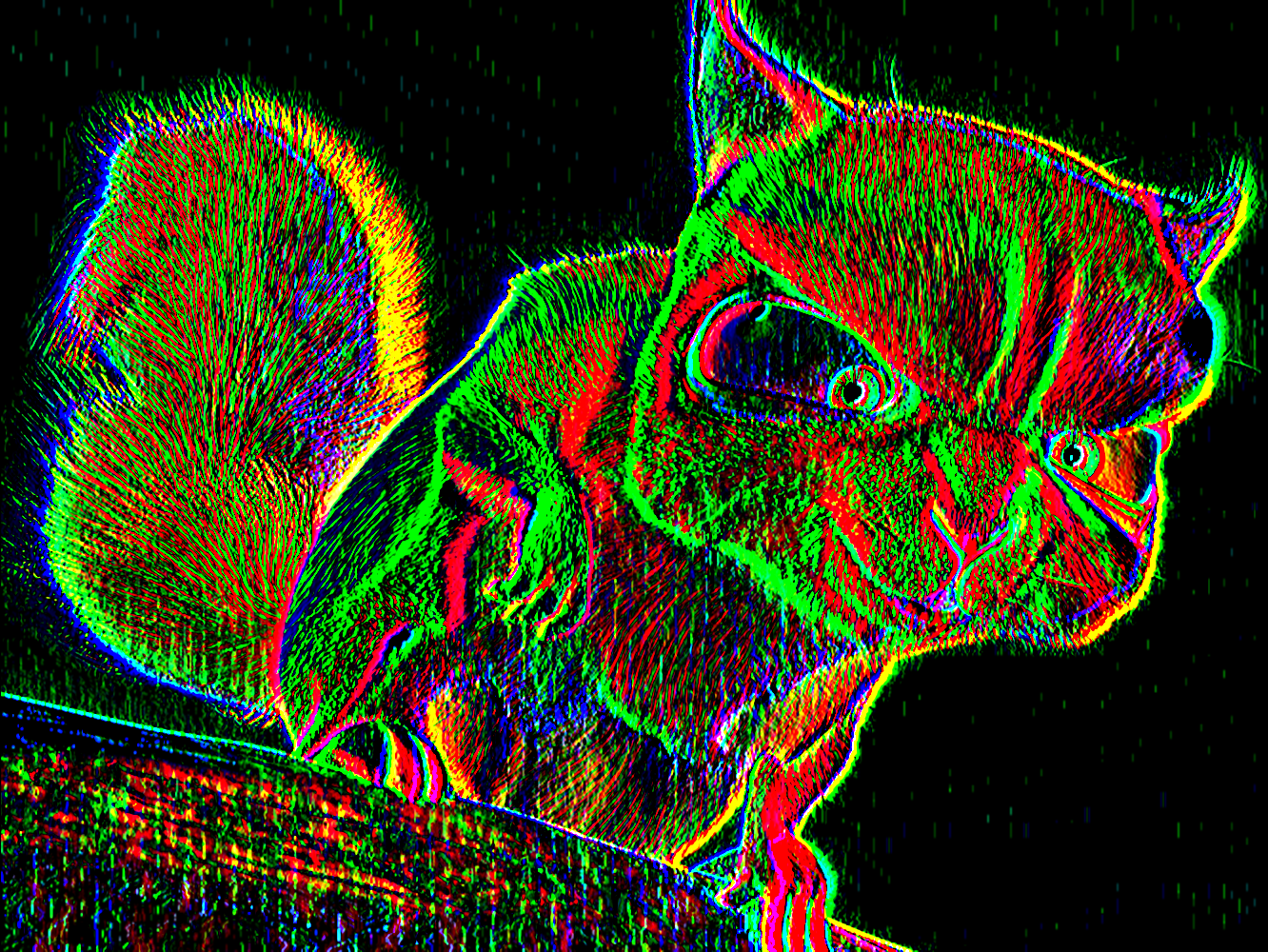}}
\end{tabular}
\vspace{-0.8em}
\\
\scalebox{.4}{\hspace{1.6em}\copyright{} Blender Foundation (\href{https://creativecommons.org/licenses/by/3.0}{CC BY 3.0})}
\begin{center}
\vspace{-0.5em}
\usebox{\mylistingop}\hfill\usebox{\mylistinghw}
\end{center}
\end{mdframed}
\caption{%
\anyhls example: The algorithm description \lst|sobel_x| is decoupled from its realization in hardware \lst|make_local_op|.
The hardware realization is a function that specifies important transformations for the exploitation of parallelism and memory architecture.
The function \lst|generate(vhls)| selects the backend for code generation, which is Vivado HLS in this case.
Ultimately, an optimized input code for HLS is generated by partially evaluating the algorithm and realization functions.
}
\label{fig:teaser}
\end{figure*}

\section{Introduction}\label{sec:intro}
\acp{FPGA} consist of a network of reconfigurable digital logic cells that can be configured to implement any combinatorial logic or sequential circuits.
This allows the design of custom application-tailored hardware.
In particular memory-intensive applications benefit from \ac{FPGA} implementations by exploiting fast on-chip memory for high throughput.
These features make \ac{FPGA} implementations orders of magnitude faster/more energy-efficient than CPU implementations in these areas.
However, \ac{FPGA} programming poses challenges to programmers unacquainted with hardware design.

\acp{FPGA} are traditionally programmed at \ac{RTL}.
This requires to model digital signals, their timing, flow between registers, as well as the operations performed on them.
\acp{HDL} such as Verilog or VHDL allow for the explicit description of arbitrary circuits but require significant coding effort and verification time.
This makes design iterations time-consuming and error-prone, even for \emph{experts}: The code needs to be rewritten for different performance or area objectives.
In recent languages such as Chisel~\cite{bachrach2012chisel}, VeriScala~\cite{liu2019scala}, and MyHDL~\cite{decaluwe2004myhdl}, programmers can create a functional description of their design but stick to the \ac{RTL}.

\ac{HLS} increases the abstraction level to an untimed high-level specification similar to imperative programming languages and automatically solves low-level design issues such as clock-level timing, register allocation, and structural pipelining~\cite{cong2011high}.
However, an \ac{HLS} code that is optimized for the synthesis of high-performance circuits is fundamentally different from a software program delivering high performance on a CPU.
This is due to the significant gap between the programming paradigms.
An \ac{HLS} compiler has to optimize the memory hierarchy of a hardware implementation and parallelize its data paths~\cite{cong2018automated}.

In order to achieve good \ac{QoR}, \ac{HLS} languages demand programmers also to specify the hardware architecture of an application instead of just its algorithm.
For this reason, \ac{HLS} languages offer hardware-specific pragmas.
This ad-hoc mix of software and hardware features makes it difficult for programmers to optimize an application.
In addition, most \ac{HLS} tools rely on their own C dialect, which prevents code portability.
For example, Xilinx Vivado~HLS~\cite{ug902} uses C++ as base language while Intel~SDK~\cite{intelsdk} (formerly Altera) uses OpenCL C.
These severe restrictions make it hard to use existing \ac{HLS} languages in a portable and modular way.

In this paper, we advocate describing \ac{FPGA} designs using functional abstractions and partial evaluation to generate optimized \ac{HLS} code.
Consider \cref{fig:teaser} for an example from image processing: %
With a functional language, we separate the description of the \lst|sobel_x| operator from its realization in hardware.
The hardware realization \lst|make_local_op| is a function that specifies the data path, the parallelization, and memory architecture.
Thus, the algorithm and hardware architecture descriptions are described by a set of higher-order functions.
A partial evaluator, ultimately, combines these functions to generate an HLS code that delivers high-performance circuit designs when compiled with HLS tools.
Since the initial descriptions are high-level, compact, and functional, they are reusable and distributable as a library.
We leverage the \anydsl compiler framework~\cite{leissa2018anydsl} to perform partial evaluation and extend it to generate input  code for \ac{HLS} tools targeting Intel and Xilinx \ac{FPGA} devices.
We claim that this approach leads to a modular and portable code other than existing \ac{HLS} approaches, and is able to produce highly efficient hardware implementations.

\noindent
In summary, this paper makes the following contributions:
\begin{itemize}
  \item
We present AnyHLS\footnote{\url{https://github.com/AnyDSL/anyhls}}, raising the abstraction level in HLS by using \emph{partial evaluation} of \emph{higher-order functions} as a core compiler technology.
It guarantees the \emph{well-typedness} of the residual program and offers considerably higher productivity than existing \ac{DSL} design techniques and C/\CC-based approaches (see \cref{sec:relwork}).
    \item
\anyhls offers unprecedented target independence, and thus portability, across different HLS tools by avoiding tool-specific pragma extensions and generating target-specific OpenCL or \CC code as input to existing HLS tools (see \cref{sec:anyhls}).
    \item
Productivity, modularity, and portability gains are demonstrated by presenting an image processing library \emph{as a case study} in \cref{sec:library}.
For this domain, we show that a competitive performance in terms of throughput and resource usage can be achieved in comparison with existing state-of-the-art DSLs (see \cref{sec:results}).
\end{itemize}

\section{Overview, Background, and Related Work}\label{sec:relwork}
\secskip
In the following, we briefly discuss prior work~(\cref{sec:relwork:qor,sec:relwork:qor,sec:relwork-abstraction}) and fundamental concepts of \anydsl (\cref{subsec:anydsl}).

\subsection{\ac{QoR} and Portability of Code in C-based HLS}\label{sec:relwork:qor}
\secskip
\ac{HLS} increases the abstraction level to an untimed high-level specification such as C/\CC or OpenCL from a fully-timed \ac{RTL}.
This eases the \emph{hardware design} problem by eliminating low-level issues such as clock-level timing, register allocation, and gate-level pipelining~\cite{pouchet2013polyhedral,cong2011high,nane2015survey}.
Modern \ac{HLS} tools are able to generate high-quality results for \ac{DSP} and datapath-oriented applications.
Several authors~(\eg \cite{cong2011high,MS09,bacon2013fpga}) have argued the following points as key to this success:
\begin{enumerate*}[label=(\roman*)]
\item advancements in \ac{RTL} design tools,
\item device-specific code generation,
\item domain-specific focus on the target applications, and
\item generating both software and hardware from the same code.
\end{enumerate*}
Modern \ac{HLS} tools such as \ac{AOCL} and Xilinx SDX offer system synthesis to map program parts to either software or hardware.
This enables software-like development for library design and verification.

There is an ongoing discussion whether C-based languages are good candidates for \ac{HLS}~\cite{edwards2006challenges,sanguinetti2006different,cong2011high,bacon2013fpga,koeplinger2018spatial}.
Yet, most commonly used HLS compilers (\eg Vivado HLS, \ac{AOCL}, Catapult, LegUp) are based on C-based languages~\cite{nane2015survey,cong2011high,ug902,intelsdk}.
The modularity and readability of C/C++ or OpenCL descriptions often conflict with best coding practices of HLS compilers~\cite{eran2019design,da2019module}.
In the hardware design context, \ac{QoR} design refers to the ratio between the performance of the circuit (latency, throughput) and design cost (circuit area, energy consumption).
A C-based \ac{HLS} code optimized for satisfactory \ac{QoR} is entirely different from a typical software program~\cite{richmond2018synthesizable,eran2019design,ozkan2017highly,da2019module,DBLP:journals/corr/abs-1805-08288}.
Thereby, the developer should express the \ac{FPGA} implementation of an application using the language abstractions of software (\ie arrays, loops to specify the memory hierarchy and hardware pipelining).
Language extensions like pragmas fill the gap for the lacking \ac{FPGA}-centric features.
However, pragmas are specific to HLS tools, and they cannot be used in a modular way because the preprocessor already resolves them (e.g., pragmas cannot be passed as function parameters).
This ad-hoc mix of software and hardware abstractions of programming languages in HLS makes optimizations hard~\cite{koeplinger2018spatial,da2019module,ozkan2017highly}.
Furthermore, the lack of standardization in \ac{HLS} languages and compilers hinders the portability of code across them.
Often, the code optimized for one HLS tool must significantly be changed to target another HLS tool even when the same FPGA design is described.
For these reasons, we believe that the next step for HLS requires an increased level of abstraction on the language side, which can reduce the need for expert knowledge. %

\subsection{Raising the Abstraction Level in HLS}\label{sec:relwork-abstraction}
\secskip

Recent work suggests raising the abstraction level in HLS by designing libraries, \acp{DSL} or source-to-source compilers to hide low-level implementation details.
This improves the modularity and reduces code duplication, but is hard to develop and maintain when well-typedness of programs are preserved.
\cite{da2019module,richmond2018synthesizable,eran2019design,ozkan2017highly} make extensive use of \CC template metaprogramming to provide libraries that are optimized for Vivado-HLS.
Generic programs can be optimized for compile-time known values using metaprogramming techniques, but it has the following drawbacks:
\begin{enumerate*}[label=(\roman*)]
\item
  The well-typedness of the generated program cannot be guaranteed in metaprogramming.
  This makes it difficult to understand error messages.
\item
  Metaprograms are hard to develop, maintain, and understand since the meta language is different from the core language (\CC core vs. \CC template language).
  For this reason, code cannot be easily moved between the core and the meta language.
\item Lambda expressions are not allowed to be used as template arguments in \CC.
\end{enumerate*}
We refer to~\cite{leissa2018anydsl} for more details.
In particular, \cite{richmond2018synthesizable,eran2019design} explain the challenges of implementing higher-order algorithms in \CC for Vivado-HLS.
OpenCL~C does not support template metaprogramming, thus forces users to use preprocessor macros for generic library design.
Therefore, libraries developed by using \CC template metaprogramming have to be rewritten completely for OpenCL~C, that is, for \ac{AOCL}.

\acp{DSL} use domain-specific knowledge to parallelize algorithms and generate low-level, optimized code~\cite{DBLP:conf/gpce/OfenbeckRSOP13}.
Programming accelerators using \acp{DSL} is thus easier, in particular for \acp{FPGA}, because the compiler performs scheduling.
A prominent example of that is the \ac{FPGA} version of Spiral~\cite{milder2012computer}.
It generates \ac{HDL} for digital signal processing applications.
In the domain of image processing, recent projects include Darkroom~\cite{hegarty2014darkroom}, Rigel~\cite{hegarty2016rigel}, and the work of \textcite{pu2017programming} based on Halide~\cite{ragan2013halide}.
\hipacc~\cite{hipaccIccad}, PolyMage~\cite{chugh2016dsl}, SODA~\cite{chi2018soda}, and RIPL~\cite{stewart2016dataflow} create image processing pipelines from a \ac{DSL}.
Rigel/Halide, PolyMage, and RIPL are declarative \acp{DSL}, whereas \hipacc is embedded into~\CC.
All of these compilers, except Rigel, generate \ac{HLS} code in order to simplify their backends.
Other examples include \textsc{Lift} that targets \acp{FPGA} via algorithmic patterns~\cite{DBLP:conf/pldi/KristienBSD19} and Tiramisu~\cite{DBLP:conf/cgo/BaghdadiRRSAZSK19} for data-parallel algorithms on dense arrays.
Tiramisu takes as input a set of scheduling commands from the user and feeds it to the polyhedral analysis of the compiler.
However, a considerable portion of these scheduling primitives remains platform-specific~\cite{FROST18}.
Spatial~\cite{koeplinger2018spatial} is a language for programming \acp{CGRA} and \acp{FPGA}.
Spatial provides language constructs to express control, memory, and interfaces of hardware implementation.

In this paper,
it is shown that the described need to raise the abstraction level in HLS may be accomplished by using recent compiler technology, in particular by exploring the concepts of \emph{partial evaluation} and \emph{high-order-functions}.
Unlike the aforementioned DSL compilers, \anyhls allows programmers to build the basic blocks and abstractions necessary for their application domain by themselves (see \cref{sec:anyhls}).
AnyHLS is thereby built on top of AnyDSL~\cite{leissa2018anydsl} (see \cref{subsec:anydsl}).
AnyDSL offers partial evaluation to enable \emph{shallow} embedding~\cite{leissa2015shallow} without the need for modifying a compiler.
This means that there is no need to change the compiler when adding support for a new application domain, since programmers can design custom control structures.
Partial evaluation specializes algorithmic variants of a program at compile-time.
Compared to metaprogramming, partial evaluation operates in a single language
and preserves the well-typedness of programs~\cite{leissa2018anydsl}.
Furthermore, different combinations of static/dynamic parameters can be instantiated from the same code.
Previously, we have shown how to abstract image border handling implementations for Intel FPGAs using AnyDSL~\cite{ozkan2018journey}.
In this paper, we present \anyhls and an image processing library to synthesize FPGA designs in a modular and abstract way for both Intel and Xilinx \acp{FPGA}.

\subsection{\anydsl Compiler Framework}\label{subsec:anydsl}
\secskip

\anydsl\footnote{\url{https://anydsl.github.io}}~\cite{leissa2015shallow,leissa2018anydsl} is a compiler framework for designing high-performance, domain-specific libraries.
It provides the imperative and functional language \impala.
Impala's syntax is inspired by Rust.
We will now briefly discuss \impala's most important features that we rely on in \anyhls.

\subsubsection{Partial Evaluation}\label{subsec:partialev}
Partial evaluation is a technique for program optimization by specialization of compile-time known values.
Assume that each input of a program $F$ is classified as either static $s$ or dynamic $d$, and values for all of the static inputs are given. Then, partial evaluation produces an optimized (residual) program $F_{s}$ such that
\begin{equation}
  [[Fs]](d) = [[F]](s,d)
\end{equation}
and running $F_{s}$ on the dynamic inputs produces the same result as running the original program $F$ on all of the inputs~\cite{jones1993partial}.
Compiler techniques such as constant propagation, loop unrolling, or inlining are examples to partial evaluation.
Typically, the user has no control when these optimizations are applied from a compiler.

\impala allows programmers to partially evaluate~\cite{DBLP:conf/rims/Futamura82} their program at compile time.
Programmers control the partial evaluator via \emph{filters}~\cite{DBLP:conf/esop/Consel88}.
These are Boolean expressions of the form \lst|@(expr)| that annotate function signatures.
\emph{Each} call site instantiates the callee's filter with the corresponding argument list.
The call is specialized when the expression evaluates to \lst|true|.
The expression \lst|?expr| yields \lst|true|, if \lst|expr| is known at compile-time;
the expression \lst|$\texttt\textdollar$expr| is never considered constant by the evaluator.
For example,
the following \lst[language=impala,basicstyle=\inlinestyle]$@(?n)$ filter will only
specialize calls to \lst[language=impala]$pow$
if \lst[language=impala]$n$ is statically known at compile-time:
\begin{lstlisting}[language=impala]
fn @(?n) pow(x: int, n: int) -> int {
  if n == 0 {
    1
  } else {
    if n %
      let y = pow(x, n / 2);
      y * y
    } else {
      x * pow(x, n - 1)
    }
  }
}
\end{lstlisting}
Thus, the calls

\noindent\begin{minipage}{0.49\columnwidth}
\begin{lstlisting}
let z = pow(x, 5);
\end{lstlisting}
\end{minipage}\hfill%
\begin{minipage}{0.49\columnwidth}
\begin{lstlisting}
let z = pow(3, 5);
\end{lstlisting}
\end{minipage}
\noindent
will result in the following equivalent sequences of instructions after specialization:

\noindent\begin{minipage}{0.49\columnwidth}
\begin{lstlisting}
let y = x * x;
let z = x * y * y;
\end{lstlisting}
\end{minipage}\hfill%
\begin{minipage}{0.49\columnwidth}
\begin{lstlisting}[showlines=true]
let z = 243;

\end{lstlisting}
\end{minipage}
As syntactic sugar, \lst|@| is available as shorthand for \lst|@(true)|.
This causes the partial evaluator to always specialize the annotated function.

FPGA implementations must be statically defined for \ac{QoR}: types, loops, functions, and interfaces must be resolved at compile-time~\cite{richmond2018synthesizable,eran2019design,ozkan2017highly}.
Partial evaluation has many advantages compared to metaprogramming as discussed
in \cref{sec:relwork-abstraction}.
Hence, Impala's partial evaluation is particularly useful to optimize HLS descriptions.

\subsubsection{Generators}\label{subsec:anydsl-highorder}

Because iteration on various domains is a common pattern, \impala provides syntactic sugar for invoking certain higher-order functions.
The loop
\begin{lstlisting}[language=impala]
for var1, ..., varn in iter(arg1, ..., argn) { /* ... */ }
\end{lstlisting}
translates to
\begin{lstlisting}[language=impala]
iter(arg1, ..., argn, |var1, ..., varn| { /* ... */ });
\end{lstlisting}
The body of the \lstinline[language=impala,basicstyle=\inlinestyle]{for} loop and the iteration variables constitute an anonymous function
\begin{lstlisting}[language=impala]
|var1, ..., varn| { /* ... */ }
\end{lstlisting}
that is passed to \lstinline[language=impala,basicstyle=\inlinestyle]{iter} as the last argument.
We call functions that are invokable like this \emph{generators}.
Domain-specific libraries implemented in \impala make busy use of these features as they allow programmers to write custom generators that take advantage of both domain knowledge and certain hardware features, as we will see in the next section.

Generators are particularly powerful in combination with partial evaluation.
Consider the following functions:
\begin{lstlisting}[mathescape=false]
type Body = fn(int) -> ();
fn @(?a & ?b) unroll(a: int, b: int, body: Body) -> () {
  if a < b { body(a); unroll(a+1, b, body) }
}
fn @ range(a: int, b: int, body: Body) -> () {
  unroll($a, b, body)
}
\end{lstlisting}
Both generators iterate from \lst|a| (inclusive) to \lst|b| (exclusive) while
invoking \lst|body| each time.
The filter \lst|unroll| tells the partial evaluator to completely unroll the recursion if both loop bounds are statically known at a particular call site.

\section{The $\text{\anyhls}$ Library}\label{sec:anyhls}
\secskip

Efficient and resource-friendly \ac{FPGA} designs require application-specific optimizations.
These optimizations and transformations are well known in the community.
For example, \textcite{DBLP:journals/corr/abs-1805-08288} discuss the key transformations of \ac{HLS} codes such as loop unrolling and pipelining.
They describe the whole hardware design from the low-level memory layout to the operator implementations with support for low-level loop transformations throughout the design.
In our setting, the programmer defines and provides these abstractions using \anydsl for a given domain in the form of a library.
We rely on partial evaluation to combine those abstractions and to remove overhead associated with them.
Ultimately, the \anydsl compiler synthesizes optimized HLS code (\CC or OpenCL C) from a given functional description of an algorithm as shown in \cref{fig:dsl_flows}.
The generated code goes to the selected \ac{HLS} tool.
This is in contrast to other domain-specific approaches like Halide-HLS~\cite{pu2017programming} or \hipacc~\cite{hipaccIccad}, which rely on domain-specific compilers to instantiate predefined templates or macros.
\hipacc makes use of two distinct libraries to synthesize algorithmic abstractions to Vivado-HLS and Intel \ac{AOCL}, while \anyhls uses the same image processing library that is described in Impala.

\subsection{HLS Code Generation}\label{subsec:hls-code-gen}
\secskip

For HLS code generation, we implemented an intrinsic named \lst|vhls| in \anyhls to emit Vivado HLS and an intrinsic named \lst|opencl| to emit \ac{AOCL}:

\noindent\begin{minipage}{0.49\linewidth}
\begin{lstlisting}
with vhls() { body() }
\end{lstlisting}
\end{minipage}\hfill%
\begin{minipage}{0.49\linewidth}
\begin{lstlisting}
with opencl() { body() }
\end{lstlisting}
\end{minipage}

\noindent With \lst|opencl| we use a grid and block size of \lst|(1, 1, 1)| to generate a single work-item kernel, as the official \ac{AOCL} documentation recommends~\cite{intelsdk}.
We extended \anydsl's OpenCL runtime by the extensions of Intel OpenCL SDK.
To provide an abstraction over both HLS backends, we create a wrapper \lst|generate| that expects a code generation function:
\begin{lstlisting}
type Backend = fn(fn() -> ()) -> ();
fn @ generate(be: Backend, body: fn() -> ()) -> () {
  with be() { body() }
}
\end{lstlisting}
Switching backends is now just a matter of passing an appropriate function to \lst|generate|:
\begin{lstlisting}
let backend = vhls; // or opencl
with generate(backend) { body() }
\end{lstlisting}

\begin{figure}
    \centering
    \scalebox{0.56}{\tikzpicturedependsonfile{figures/src/dsl_flows.tikz.tex}

\begin{tikzpicture}
    \usetikzlibrary{%
      arrows,%
      shapes.misc,%
      shapes.arrows,%
      chains,%
      graphs,
      matrix,%
      positioning,%
      scopes,%
      decorations.pathmorphing,%
      backgrounds, fit,
      shadows,%
      calc
    }

    \def\nodewidth{4em}
    \def\nodeheight{5ex}

    \newlength\nodedist
    \newlength\danch
    \newlength\dsplit
    \setlength\nodedist{5em}
    \setlength\danch{11em}
    \setlength\dsplit{2em}

    \definecolor{runtimecolor}{RGB}{255, 152, 0}
    \definecolor{backendcolor}{RGB}{3, 169, 244}

    \tikzstyle{nonterminal} = [%
      rectangle,rounded corners,
      node distance=\nodedist,%
      minimum width=\nodewidth,%
      minimum height=\nodeheight,%
      thick, draw=black,
      font=\itshape,
      align=center]%

    \tikzstyle{backend} = [%
      rectangle,
      node distance=\nodedist,%
      minimum width=\nodewidth,%
      minimum height=\nodeheight,%
      thick, draw=black,
      font=\itshape,
      fill=gray!30,
      align=center]%

   \tikzstyle{textbox} = [%
     rectangle, draw=none,
     node distance=\nodedist,%
     font=\itshape,
     align=center]%

    \tikzstyle{backendwhite} = [%
      rectangle,
      node distance=0.5*\nodedist,%
      minimum width=\nodewidth,%
      minimum height=\nodeheight,%
      thick, draw=black,
      fill=gray!60,
      font=\itshape,
      align=center]%

    \tikzstyle{library} = [%
      rectangle,
      node distance=\nodedist,%
      minimum width=\nodewidth,%
      minimum height=\nodeheight,%
      thick, draw=black,
      font=\itshape,
      fill=gray!60,
      align=center]%

    \tikzstyle{board} = [%
      align=center,%
      minimum width=\nodewidth,%
      node distance=\nodedist]

    \tikzstyle{cord} = [%
      coordinate,%
      node distance=\nodedist,%
      align=center]%

    \tikzstyle{arrw} = [%
      black,
      semithick,
      -triangle 45]

    \node (halide-abs) [nonterminal] {halide-app.cpp};
    \node (hipacc-abs) [nonterminal, right = \danch*0.5 of halide-abs] {hipacc-app.cpp};
    \node (anydsl-abs) [nonterminal, right = \danch*0.7 of hipacc-abs]{anyhsl-app.impala};

    \newlength\dyTwo
    \pgfmathparse{0.5 * \nodedist}
    \setlength\dyTwo{\pgfmathresult pt}
    \node (hipacc-comp-lab) [textbox, below = 0.6*\dsplit of hipacc-abs] {Hipacc compiler};
    \node (hipacc-comp-lab-coord) [coord, below = 0.75*\dsplit of hipacc-abs] {};
    \node (hipacc-below)    [cord, below = 2*\dsplit of hipacc-abs] {};

    \node (anydsl-compiler) [backendwhite, below = 0.75*\dsplit of anydsl-abs] {\vspace{2pt} AnyDSL compiler\\ + \\ (partial evaluator) \vspace{2pt}};
    \node (halide-back-hls) [backend,     below of = halide-abs] {Vivado \\ backend};
    \node (hipacc-back-hls) [backend,     left  =     \dsplit of hipacc-below] {Vivado \\ backend};
    \node (hipacc-back-aoc) [backend,     right = 0.6*\dsplit of hipacc-below] {AOCL \\ backend};

    \node (halide-comp-lab) [textbox, below = 0.6*\dsplit of halide-abs] {Halide compiler};
    \node (halide-comp-lab-coord) [coord, below = 0.75*\dsplit of halide-abs] {};

    \begin{pgfonlayer}{background}
      \node[backendwhite] (hipacc-comp) [fit = (hipacc-back-hls) (hipacc-back-aoc) (hipacc-comp-lab-coord)] {};
      \node[backendwhite, minimum width=1.75*\nodewidth] (halide-comp) [fit = (halide-back-hls) (halide-comp-lab-coord)] {};
    \end{pgfonlayer}

    \node (anydsl-back)     [nonterminal, right = 0.5*\dsplit of anydsl-compiler] {Image \\ Processing \\ Lib.impala};
    \node (anydsl-back-below) [cord, below = 0.7*\nodedist of anydsl-compiler] {};
    \node (anydsl-back-below-above) [cord, above = 0.25*\nodedist of anydsl-back-below] {};

    \node (hipacc-hls) [nonterminal, below of = hipacc-back-hls] {VHLS-code.cpp};
    \node (hipacc-aoc) [nonterminal, below of = hipacc-back-aoc] {AOCL-code.cl};
    \node (halide-hls) [nonterminal, below of = halide-back-hls] {VHLS-code.cpp};
    \node (anydsl-hls) [nonterminal, left  = 0.5em of anydsl-back-below] {VHLS-code.cpp};
    \node (anydsl-aoc) [nonterminal, right = 0.5em of anydsl-back-below] {AOCL-code.cl};

    \newlength\dyFourLib
    \pgfmathparse{0.6*\nodedist}
    \setlength\dyFourLib{\pgfmathresult pt}
    \node (anydsl-hls-below) [cord, below = \dyFourLib of anydsl-hls] {};
    \node (anydsl-aoc-below) [cord, below = \dyFourLib of anydsl-aoc] {};
    \node (hipacc-hls-below) [cord, below = \dyFourLib of hipacc-hls] {};
    \node (hipacc-aoc-below) [cord, below = \dyFourLib of hipacc-aoc] {};
    \node (halide-hls-below) [cord, below = \dyFourLib of halide-hls] {};

    \node (hipacc-hls-lib) [library, left  = -2ex of hipacc-hls-below ] {template \\ library};
    \node (hipacc-aoc-lib) [library, right = -2ex of hipacc-aoc-below] {template \\ library};
    \node (halide-hls-lib) [library, left  = -2ex of halide-hls-below ] {template \\ library};

    \newlength\dyFourEnd
    \pgfmathparse{0.4*\nodedist}
    \setlength\dyFourEnd{\pgfmathresult pt}
    \node (anydsl-xilinx-above) [cord, below = \dyFourEnd of anydsl-hls-below] {};
    \node (anydsl-altera-above) [cord, below = \dyFourEnd of anydsl-aoc-below] {};
    \node (hipacc-xilinx-above) [cord, below = \dyFourEnd of hipacc-hls-below] {};
    \node (hipacc-altera-above) [cord, below = \dyFourEnd of hipacc-aoc-below] {};
    \node (halide-xilinx-above) [cord, below = \dyFourEnd of halide-hls-below] {};

    \newlength\dyFour
    \pgfmathparse{\dyFourLib + \dyFourEnd}
    \setlength\dyFour{\pgfmathresult pt}

    \newlength\dyFive
    \pgfmathparse{0.3*\nodedist}
    \setlength\dyFive{\pgfmathresult pt}
    \node(hipacc-vhls)  [backendwhite, below = \dyFive of hipacc-xilinx-above, align=center] {VHLS};
    \node(hipacc-aocl)  [backendwhite, below = \dyFive of hipacc-altera-above, align=center] {AOCL};
    \node(anydsl-vhls)  [backendwhite, below = (\dyFive) of anydsl-xilinx-above, align=center] {VHLS};%
    \node(anydsl-aocl)  [backendwhite, below = (\dyFive) of anydsl-altera-above, align=center] {AOCL};%
    \node(halide-vhls)  [backendwhite, below = \dyFive of halide-xilinx-above, align=center] {VHLS};%

    \node(hipacc-xilinx)  [nonterminal, below = \dyFive of hipacc-vhls, align=center] {XILINX \\ FPGA};
    \node(hipacc-altera)  [nonterminal, below = \dyFive of hipacc-aocl, align=center] {INTEL  \\ FPGA};
    \node(anydsl-xilinx)  [nonterminal, below = \dyFive of anydsl-vhls, align=center] {XILINX \\ FPGA };%
    \node(anydsl-altera)  [nonterminal, below = \dyFive of anydsl-aocl, align=center] {INTEL  \\ FPGA };%
    \node(halide-xilinx)  [nonterminal, below = \dyFive of halide-vhls, align=center] {XILINX \\ FPGA };%

    {
        \draw[arrw] (anydsl-abs) -- (anydsl-compiler);
        \draw[arrw] (anydsl-back) -- (anydsl-compiler);
        \draw[arrw] (anydsl-compiler.south west)++(+.6,0) |- ++(0,-.15*\nodedist) -| (anydsl-hls);
        \draw[arrw] (anydsl-compiler.south east)++(-.6,0) |- ++(0,-.15*\nodedist) -| (anydsl-aoc);
        \draw[arrw] (anydsl-hls)  -- (anydsl-vhls);
        \draw[arrw] (anydsl-aoc) -- (anydsl-aocl);

        \draw[arrw] (hipacc-abs) -- (hipacc-comp);
        \draw[arrw] (hipacc-back-hls)  -- (hipacc-hls);
        \draw[arrw] (hipacc-back-aoc) -- (hipacc-aoc);

        \draw[arrw] (halide-abs) -- (halide-comp);
        \draw[arrw] (halide-comp) -- (halide-hls);

        \draw[arrw] (hipacc-hls.south)++(+.5,0) |- ++(0,-\dyFour) -| ($(hipacc-vhls.north) + (+0.2,0)$);
        \draw[arrw] (hipacc-hls-lib) |- ++(0.2,-\dyFourEnd) -| ($(hipacc-vhls.north) + (-0.2,0)$);
        \draw[arrw] (hipacc-aoc.south)++(-.5,0) |- ++(0,-\dyFour) -| ($(hipacc-aocl.north) + (-0.2,0)$);
        \draw[arrw] (hipacc-aoc-lib) |- ++(-0.2,-\dyFourEnd) -| ($(hipacc-aocl.north) + (+0.2,0)$);
        \draw[arrw] (halide-hls.south)++(+.5,0) |- ++(0,-\dyFour) -| ($(halide-vhls.north) + (+0.2,0)$);
        \draw[arrw] (halide-hls-lib) |- ++(0.2,-\dyFourEnd) -| ($(halide-vhls.north)+ (-0.2,0)$);

        \draw[arrw] (halide-vhls) -- (halide-xilinx);
        \draw[arrw] (hipacc-vhls) -- (hipacc-xilinx);
        \draw[arrw] (hipacc-aocl) -- (hipacc-altera);
        \draw[arrw] (anydsl-vhls) -- (anydsl-xilinx);
        \draw[arrw] (anydsl-aocl) -- (anydsl-altera);
    }

    \tikzstyle{outerbox} = [draw=black, dashed, rounded corners, rectangle]
    \begin{pgfonlayer}{background}
      \node[outerbox] [fit = (anydsl-abs) (anydsl-back) (anydsl-hls) (anydsl-aoc)
                             (anydsl-xilinx) (anydsl-altera) ] {};
    \end{pgfonlayer}
    \begin{pgfonlayer}{background}
      \node[outerbox] [fit = (hipacc-abs) (hipacc-back-hls) (hipacc-back-aoc)
                             (hipacc-hls) (hipacc-aoc) (hipacc-aoc-lib)
                             (hipacc-hls-lib) (hipacc-xilinx) (hipacc-altera)] {};
    \end{pgfonlayer}
    \begin{pgfonlayer}{background}
      \node[outerbox] [fit = (halide-abs) (halide-back-hls) (halide-hls)
                             (halide-hls-lib) (halide-xilinx)] {};
    \end{pgfonlayer}
\end{tikzpicture}}
    \caption{
        FPGA code generation flows for Halide, \hipacc, and \anyhls (from left to right).
        VHLS and AOCL are used as acronyms for Vivado HLS and Intel FPGA SDK for OpenCL, respectively.
        Halide and \hipacc rely on domain-specific compilers for image processing that instantiate template libraries.
        \anyhls allows defining all abstractions for a domain in a language called \impala and relies on partial evaluation for code specialization.
        This ensures maintainability and extensibility of the provided domain-specific library---for image processing in this example.
    }
    \label{fig:dsl_flows}
\end{figure}
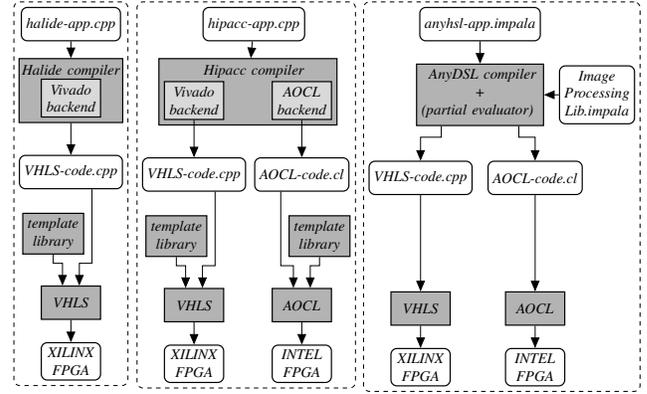

\subsection{Building Abstractions for FPGA Designs}\label{sec:overview}
\secskip
In the following, we present abstractions for the key transformations and design patterns that are common in FPGA design.
These include (a) important loop transformations, (b) control flow and data flow descriptions such as reductions, \acp{FSM} and (d) the explicit utilization of different memory types.
Approaches like Spatial~\cite{koeplinger2018spatial} expose these patterns within the language---new patterns require dedicated support from the compiler.
Hence, these languages and compilers are restricted to a specialized application domain they have been designed for.
In \anyhls, \impala's functional language and partial evaluation allow us to design the abstractions needed for FPGA synthesis in the form of a library.
New patterns can be added to the library without dedicated support from the compiler.
This makes \anyhls easier to extend compared to the approaches mentioned afore.

\subsubsection{Loop Transformations}
\label{subsec:combining_transformations}

\CC compilers usually provide certain preprocessor directives that perform particular code transformations.
A common feature is to unroll loops (see left-hand side):

\noindent\begin{minipage}{0.49\linewidth}
\begin{lstlisting}[language=C++,otherkeywords={\#pragma}]
for (int i=0; i<N/W; ++i) {
  for (int w=0; w<W; ++w) {
    #pragma unroll
    body(i*W + w);
  }
}
\end{lstlisting}
\end{minipage}\hfill%
\begin{minipage}{0.49\linewidth}
\begin{lstlisting}
for i in range(0, N/W) {
  for w in unroll(0, W) {

    body(i*W + w);
  }
}
\end{lstlisting}
\end{minipage}%

\begin{figure}
  \centering
  \scalebox{0.65}{\tikzpicturedependsonfile{figures/src/anyhls_lib.tikz.tex}
\tikzpicturedependsonfile{figures/src/anyhls_parallel_processing.tikz.tex}
\begin{tikzpicture}
    \matrix[column sep=1mm, row sep=1mm] {
        \node (pe10) { \usebox\anyhlsPEOne    }; &
        \node (pe11) { \usebox\anyhlsPEOneOne }; &
        \node (pe20) { \usebox\anyhlsPETwo    }; &
        \node (pe22) { \usebox\anyhlsPETwoTwo }; \\
    };
\end{tikzpicture}}
  \caption{Parallel processing}
  \label{fig:anyhls:parallel:processing}
\end{figure}
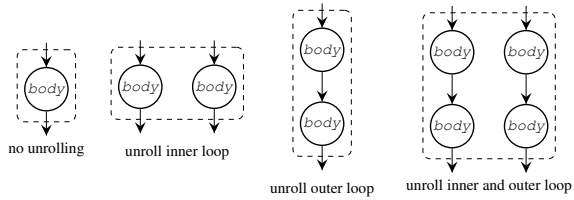

\noindent
Such \mbox{\lst|pragma|s} are built into the compiler.
The \impala version (shown at right) uses generators that are entirely implemented as a library.
Partial evaluation optimizes Impala's \lst{range} and \lst{unroll} abstractions as well as the input body function according to their static inputs, \ie \lst{N}, \lst{W}.
The residual program consists of the consecutive $body$ function according to the value of the \lst{W} as shown in \cref{fig:anyhls:parallel:processing}.
This generates a concise and clean code for the target HLS compiler, which is drastically different from using a pragma.

Generators, unlike \CC \lst[language=c++]{pragma}s, are first-class citizens of the Impala language.
This allows programmers to implement sophisticated loop transformations.
For example, the following function \lst|tile| returns a new generator.
It instantiates a tiled loop nest of the specified tile \lst|size| with the \mbox{\lst|Loop|s} \lst|inner| and \lst|outer|:
\begin{lstlisting}
type Loop = fn(int, int, fn(int) -> ()) -> ();
fn @ tile(size: int, inner: Loop, outer: Loop) -> Loop {
  @|beg, end, body| outer(0, (end-beg)/size,
    |i| inner(i*size + beg, (i+1)*size + end, |j| body))
}

let schedule = tile(W, unroll, range);
for i in schedule(0, N) {
  body(i)
}
\end{lstlisting}
Passing \lst|W| for the tiling \lst|size|, \lst|unroll| for the inner loop, and \lst|range| for the outer loop yields a generator that is identical to the loop nest at the beginning of this paragraph.
With this design, we can reuse or explore iteration techniques without touching the actual body of a \lst|for| loop.
For example, consider the processing options for a two-dimensional loop nest as shown in \cref{fig:anyhls:parallel:processing}:
When just passing \lst|range| as \lst|inner| and \lst|outer| loop, the partial evaluator will keep the loop nest and, hence, not unroll \lst|body| and instantiate it only once.
Unrolling the inner loop replicates \lst|body| and increases the bandwidth requirements accordingly.
Unrolling the outer loop also replicates \lst|body|, but in a way that benefits data reuse from the temporal locality of an iterative algorithm.
Unrolling both loops replicate \lst|body| for increased bandwidth and data reuse for the temporal locality.

C/\CC-based \ac{HLS} solutions often use a \lst|pragma| to mark a loop amenable for pipelining.
This means parallel execution of the loop iterations in hardware.
For example, the following code on the left uses an initiation interval~(\lst|II|) of~\lst|3|:

\noindent\begin{minipage}{0.47\columnwidth}
\begin{lstlisting}[language=C++,otherkeywords={\#pragma}]
for (int i=0; i<N; ++i) {
  #pragma HLS pipeline II=3
  body(i);
}
\end{lstlisting}
\end{minipage}\hfill%
\begin{minipage}{0.51\columnwidth}
\begin{lstlisting}
let II = 3;
for i in pipeline(II, 0, N) {
  body(i)
}
\end{lstlisting}
\end{minipage}

\noindent
Instead of a pragma (on the left), AnyHLS uses the intrinsic generator \lst|pipeline| (on the right).
Unlike the above loop abstractions (\eg unroll), Impala emits a tool-specific pragma for the \lst{pipeline} abstraction.
This provides portability across different HLS tools.
Furthermore, it allows the programmer to invoke and pass around \lst|pipeline|---just like any other generator.

\subsubsection{Reductions}
\label{sec:reductions}

Reductions are useful in many contexts.
The following function takes an array of values, a range within, and an operator:
\begin{lstlisting}
type T = int;
fn @(?beg & ?end) reduce(beg: int, end: int, input: &[T],
                         op: fn(T, T) -> T) -> T {
  let n = end - beg;
  if n == 1 {
    input(beg)
  } else {
    let m = (end + beg) / 2;
    let a = reduce(beg, m, input, op);
    let b = reduce(m, end, input, op);
    op(a, b)
  }
}
\end{lstlisting}
In the above filter, the recursion will be completely unfolded if the range is statically known.
Thus,
\begin{lstlisting}
reduce(0, 4, [a, b, c, d], |x, y| x + y)
\end{lstlisting}
yields: \lst|(a + b) + (c + d)|.

\subsubsection{Finite State Machines}\label{sec:fsm}
AnyHLS models computations that depend not only on the inputs but also on an internal state with an \ac{FSM}.
To define an \ac{FSM}, programmers need to specify states and a transition function that determines when to change the current state based on the machine's input.
This is especially beneficial for modeling control flow.
To describe an \ac{FSM} in \impala, we start by introducing types to represent the states and the machine itself:
\begin{lstlisting}
type State = int;
struct FSM {
  add: fn(State, fn() -> (), fn() -> State) -> (),
  run: fn(State) -> ()
}
\end{lstlisting}
An object of type \lst|FSM| provides two operations: adding one state with \lst|add| or \mbox{\lst|run|ning} the computation.
The \lst|add| method takes the name of the state, an action to be performed for this state, and a transition function associated with this state.
Once all states are added, the programmer \mbox{\lst|run|s} the machine by passing the initial state as an input parameter.
The following example adds~\lst|1| to every element of an array:
\begin{lstlisting}
let buf = /*...*/;
let mut (idx, pixel) = (0, 0);
let fsm = make_fsm();
fsm.add(Read, || pixel = buf(idx),
              || if idx>=len { Exit } else { Compute });
fsm.add(Compute, || pixel += 1,         || Write);
fsm.add(Write,   || buf(idx++) = pixel, || Read );
fsm.run(Read);
\end{lstlisting}
Similar the other abstractions introduced in this section, the constructor for an \ac{FSM} is not a built-in function of the compiler but a regular \impala function.
In some cases, we want to execute the \lst|FSM| in a pipelined way.
For this scenario, we add a second method \lst|run_pipelined|.
As all the methods, \eg \lst|make_fsm|, \lst|add|, \lst|run|, are annotated for partial evaluation (by \lst|@|), input functions to these methods will be optimized according to their static inputs.
Ultimately, AnyHLS will emit the states of an \lst|FSM| as part of a loop according to the selected \lst|run| method.

\subsubsection{Memory Types and Memory Abstractions}\label{subsec:anyhls-core-mem}

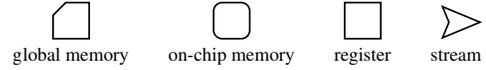
\begin{figure}[!t]
  \centering
  \scalebox{0.8}{\tikzpicturedependsonfile{figures/src/anyhls_lib.tikz.tex}
\tikzpicturedependsonfile{figures/src/anyhls_memory_types.tikz.tex}
\begin{tikzpicture}
    \matrix[column sep=1mm, row sep=1mm] {
        \node (global-mem) { \usebox\anyhlsGlobal }; &
        \node (onchip-mem) { \usebox\anyhlsOnchip }; &
        \node (reg-mem)    { \usebox\anyhlsReg    }; &
        \node (stream-mem) { \usebox\anyhlsStream }; \\
    };
\end{tikzpicture}}
  \caption{Memory types provided for FPGA design}
  \label{fig:anyhls:memory:types}
\end{figure}

FPGAs have different memory types of varying sizes and access properties.
Impala supports four memory types specific to hardware design (see \cref{fig:anyhls:memory:types}): global memory, on-chip memory, registers, and streams.
Global memory (typically DRAM) is allocated on the host using our runtime and accessed through regular pointers.
On-chip memory (\eg BRAM or M10K/M20K) for the FPGA is allocated using the \lst|reserve_onchip| compiler intrinsic.
Memory accesses using the pointer returned by this intrinsic will map to on-chip memory.
Standard variables are mapped to registers, and a specific \lst|stream| type is available to allow for the communication between FPGA kernels.
Memory-wise, a \lst|stream| is mapped to registers or on-chip memory by the HLS tools.
These FPGA-specific memory types in Impala will be mapped to their corresponding tool-specific declarations in the residual program (on-chip memory will be defined as local memory for \ac{AOCL}  whereas it will be defined as an array in Vivado HLS).

\paragraph{Memory partitioning}
an array partitioning pragma must be defined as follows to implement a C array with hardware registers using Vivado HLS~\cite{ug902}:
\begin{lstlisting}[
  language=C++,
  label={lst:partition-hls},
  caption={A typical way of partitioning an array by using pragmas in existing HLS tools.}
]
typedef int T;
T Regs1D[size];
#pragma HLS variable=Regs1D array_partition dim=0
\end{lstlisting}
\bigskip
\noindent
Other HLS tools offer similar pragmas for the same task.
Instead, \anyhls provides a more concise description of a register array without using any tool-specific pragma by the recursive declaration of registers as follows:
\begin{lstlisting}[
  label={lst:partition-impala},
  caption={Recursive description of a register array using partial evalution instead of declaring an array and partitioning it by HLS pragmas.}
]
type T = int;
struct Regs1D {
  read:  fn(int) -> T,
  write: fn(int, T) -> (),
  size:  int
}
fn @ make_regs1d(size: int) -> Regs1D {
  if size == 0 {
    Regs1D {
      read:  @|_| 0,
      write: @|_, _| (),
      size:  size
    }
  } else {
    let mut reg: T;
    let others = make_regs1d(size - 1);
    Regs1D {
      read:  @|i|    if i+1 == size { reg }
                      else { others.read(i) },
      write: @|i, v| if i+1 == size { reg = v }
                      else { others.write(i, v) },
      size:  size
    }
  }
}
\end{lstlisting}

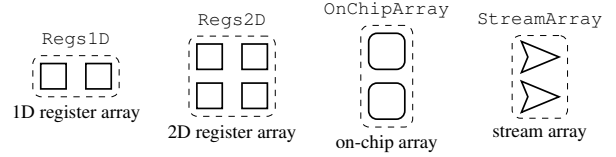
\begin{figure}[!t]
  \centering
  \scalebox{0.8}{\hspace{-2em}\tikzpicturedependsonfile{figures/src/anyhls_lib.tikz.tex}
\tikzpicturedependsonfile{figures/src/anyhls_memory_abstractions.tikz.tex}
\begin{tikzpicture}
    \matrix[column sep=-1mm, row sep=1mm] {
        \node (regs-array-one) { \usebox\anyhlsRegsOneD   }; &
        \node (regs-array-two) { \usebox\anyhlsRegsTwoD   }; &
        \node (onchip-mem-two) { \usebox\anyhlsOnchipTwoD }; &
        \node (stream-mem-two) { \usebox\anyhlsStreamTwoD }; \\
    };
\end{tikzpicture}}
  \caption{Memory abstractions}
  \label{fig:anyhls:memory:abstractions}
\end{figure}

\bigskip
When the \lst|size| is not zero, each recursive call to this function allocates a register variable named \lst|reg|, and creates a smaller register array with one element less named \lst|others|.
The \lst|read| and \lst|write| functions test if the index \lst|i| is equal to the index of the current register.
In the case of a match, the current register is used.
Otherwise, the search continues in the smaller array.
The generator (\lst|make_regs1d|) returns an Impala variable that can be read and written by index values (\lst|regs| in the following code), similar to C arrays.
\begin{lstlisting}
let regs = make_regs1d(size);
\end{lstlisting}
However, it defines \lst|size| number of registers in the residual program instead of declaring an array and partitioning it by tool-specific pragmas as in \cref{lst:partition-hls}.
The generated code does not contain any compiler directives; hence it can be used for different HLS tools (\eg Vivado HLS, \ac{AOCL}).
Since we annotated \lst|make_regs1d|, \lst|read|, and \lst|write| for partial evaluation, any call to these functions will be inlined recursively.
This means that the search to find the register to read to or write from will be performed at compile time. %
These registers will be optimized by the AnyDSL compiler, just like any other variables:
unnecessary assignments will be avoided, and a clean HLS code will be generated.

Correspondingly, \anyhls provides generators (similar to \cref{lst:partition-impala}) for one and two-dimensional arrays of on-chip memory (\eg line buffers in \cref{sec:library}), global memory, and streams (as illustrated in \cref{fig:anyhls:memory:abstractions}) instead of using memory partitioning pragmas encouraged in existing HLS tools (as in \cref{lst:partition-hls}).

\section{A Library for Image Processing on FPGA}\label{sec:library}
\secskip
AnyHLS allows for defining domain-specific abstractions and optimizations that are used and applied prior to generating customized input to existing HLS tools.
In this section, we introduce a library that is developed to support \ac{HLS} for the domain of image processing applications.
It is based on the fundamental abstractions introduced in \cref{sec:overview}.
Our low-level implementation is similar to existing domain-specific languages targeting \acp{FPGA}~\cite{hegarty2016rigel,hipaccIccad}.
For this reason, we focus on the interface of our abstractions as seen by the programmer.

We design applications by decoupling their algorithmic description from their schedule and memory operations.
For instance, typical image operators, such as the following Sobel filter, just resort to the \lst|make_local_op| generator.
Similarly, we implement a point operator for RGB-to-gray color conversion as follows (\cref{lst:app-ex}):
\begin{lstlisting}[
  label={lst:app-ex},
  caption={Sobel filter and RGB-to-gray color conversion as example applications described by using our library. }
  ]
fn sobel_edge(output: &mut [T], input: &[T]) -> () {
  let img = make_raw_mem2d(width, height,  input);
  let dx  = make_raw_mem2d(width, height, output);
  let sobel_extents = extents(1, 1); // for 3x3 filter
  let operator = make_local_op(4,    // vector factor
    sobel_operator_x, sobel_extents, mirror, mirror);
  with generate(hls) { operator(img, dx); }
}

fn rgb2gray(output: &mut [T], input: &[T]) -> () {
  let img  = make_raw_img(width, height,  input);
  let gray = make_raw_img(width, height, output);
  let operator = make_point_op(@ |pix| {
    let r = pix & 0xFF;
    let g = (pix >>  8) & 0xFF;
    let b = (pix >> 16) & 0xFF;
    (r + g + b) / 3
  });
  with generate(hls) { operator(img, gray); }
}
\end{lstlisting}
\bigskip

\noindent
The image data structure is opaque.
The target platform mapping determines its layout.
\anyhls provides common border handling functions as well as point and global operators such as reductions (see \cref{sec:reductions}).
These operators are composable to allow for more sophisticated ones.

\subsection{Vectorization}\label{sec:library:vectorization}
\secskip
Image processing applications consist of loops that possess a very high degree of spatial parallelism.
This should be exploited to reach the bandwidth speed of memory technologies.
A resource-efficient approach, so-called \emph{vectorization} or \emph{loop coarsening}, is to aggregate the input pixels to vectors and process multiple input data at the same time to calculate multiple output pixels in parallel~\cite{schmid2015loop,ASAP17,stitt2018scalable}.
This replicates only the arithmetic operations applied to data (so-called datapath) instead of the whole accelerator, similar to \ac{SIMD} architectures.
Vectorization requires a control structure specialized to a considered hardware design.
We support the automatic vectorization of an application by a given factor \lst|v| when using our image processing library.
In particular, our library use the vectorization techniques proposed in~\cite{ASAP17}.
For example, the \lst|make_local_op| function has an additional parameter to specify the desired vectorization and will propagate this information to the functions it uses internally: \lst|make_local_op(op, v)|.
For brevity, we omit the parameter for the vectorization factor for the remaining abstractions in this section.

\subsection{Memory Abstractions for Image Processing}
\secskip

\subsubsection{Memory Accessor}
In order to optimize memory access and encapsulate the contained memory type (on-chip memory, etc.) into a data structure, we decouple the data transfer from the data use via the following memory abstractions:

\noindent\begin{minipage}{0.44\linewidth}
\begin{lstlisting}
struct Mem1D {
  read:   fn(int) -> T,
  write:  fn(int, T)->(),
  update: fn(int) -> (),
  size:   int
}
\end{lstlisting}
\end{minipage}\hfill%
\begin{minipage}{0.54\linewidth}
\begin{lstlisting}
struct Mem2D {
  read:   fn(int, int) -> T,
  write:  fn(int, int, T)->(),
  update: fn(int, int) -> (),
  width:  int, height: int
}
\end{lstlisting}
\end{minipage}
Similar to hardware design practices, these memory abstractions require the memory address to be \mbox{\lst{update}d} before the \lst{read}/\lst{write} operations.
The \lst{update} function transfers data from/to the encapsulated memory to/from staging registers using vector data types.
Then, the \lst{read}/\lst{write} functions access an element of the vector.
This increases data reuse and DRAM-to-on-chip memory bandwidth~\cite{choi2016quantitative}.

\subsubsection{Stream Processing}\label{sec:library:stream}
Inter-kernel dependencies of an algorithm should be accessed on-the-fly in combination with fine-granular communication in order to pipeline the full implementation with a fixed throughput.
That is, as soon as a block produces one data, the next block consumes it.
In the best case, this requires only a single register of a small buffer instead of reading/writing to temporary images:
\newline

\noindent\makebox[\linewidth]{%
\scalebox{0.63}{\tikzpicturedependsonfile{figures/src/streams.tikz.tex}

\begin{tikzpicture}%
  \def\dx{20} \def\dy{30} %

  \tikzstyle{textbox} = [%
        rectangle, draw=none, font=\small,
    node distance=\nodedist,%
    align=center]%

  \tikzstyle{cord} = [%
    coordinate,%
    node distance=\nodedist,%
    align=center]%

  \tikzset{arrw/.style={decoration={markings,mark=at position 1 with %
    {\arrow[scale=2,>=stealth]{>}}},postaction={decorate}}}

  \tikzset{nonterminal/.append style={text height=1.5ex,text depth=.25ex}}
  \tikzset{
      global/.style={chamfered rectangle, chamfered rectangle corners=north west, minimum size=6mm, draw=black, font=\itshape},
      nonterminal/.style={rectangle, rounded corners, minimum size=6mm, thick, draw=black, font=\itshape},
      register/.style={rectangle, minimum size=6mm, thick, draw=black, font=\itshape},
      stream/.style={dart, minimum size=3mm, thick, draw=black, font=\itshape},
      splitStream/.style={diamond, double, minimum size=6mm, thick, draw=black, font=\itshape},
      pointop/.style ={draw=black, chamfered rectangle, double},
      localop/.style ={draw=black, rectangle, inner ysep=8pt, double},
      terminal/.style={rounded rectangle, minimum size=6mm, thick,draw=black!50, font=\ttfamily},
      skip loop/.style={to path={-- ++(0,#1) -| (\tikztotarget)}}
  }
  \tikzstyle{outerbox} = [draw=black, dashed, rounded corners, rectangle]

  \tikzstyle{shared} = [rectangle, rounded corners, minimum height=14pt, minimum width=12pt,
  inner sep=0pt, draw=black, thick, node distance = \dx]

  \node [localop] (opC) {\Large Kernel2};
  \node [stream, left = \dx pt of opC] (bufL-mem)  {};
  \node [stream, right= \dx pt of opC] (bufR-mem)  {};
  \node [localop,  left = \dx/1.5 pt of bufL-mem] (opL) {\Large Kernel1};
  \node [localop,  right= \dx pt of bufR-mem] (opR) {\Large Kernel3};

  \node [global, left = \dx/1.5 pt of opL] (global-in-mem)  {\phantom{Mem1D}};
  \node [global, right = \dx/1.5 pt of opR] (global-out-mem) {\phantom{Mem1D}};

  \begin{pgfonlayer}{background}
    \node[outerbox] (global-in) [fit = (global-in-mem)] {};
    \node[outerbox] (global-out) [fit = (global-out-mem)] {};
    \node[outerbox] (bufL) [fit = (bufL-mem)] {};
    \node[outerbox] (bufR) [fit = (bufR-mem)] {};
  \end{pgfonlayer}
  \node (global-in-label)[textbox, above = 0.01cm of global-in-mem] {Mem1D};
  \node (global-out-label)[textbox, above = 0.01cm of global-out-mem] {Mem1D};
  \node (bufL-label)[textbox, above = 0.01cm of bufL] {Mem1D};
  \node (bufR-label)[textbox, above = 0.01cm of bufR] {Mem1D};

  \draw[arrw] (global-in) -- (opL);
  \draw[arrw] (opL) -- (bufL);
  \draw[arrw] (bufL) -- (opC);
  \draw[arrw] (opC) -- (bufR);
  \draw[arrw] (bufR) -- (opR);
  \draw[arrw] (opR) -- (global-out);

\end{tikzpicture}}%
}
\vspace{0.01em}

\noindent
We define a stream between two kernels as follows:
\begin{lstlisting}
fn make_mem_from_stream(size: int, data: stream) -> Mem1D;
\end{lstlisting}

\subsubsection{Line Buffers}\label{sec:library:linebuffer}
Storing an entire image to on-chip memory before execution is not feasible since on-chip memory blocks are limited in \acp{FPGA}.
On the other hand, feeding the data on demand from main memory is extremely slow.
Still, it is possible to leverage fast on-chip memory by using it as FIFO buffers containing only the necessary lines of the input images ($W$ pixels per line).

\noindent\makebox[\linewidth]{%
\scalebox{0.85}{\usebox\anyhlsLineBufferDetail}%
}

\noindent
This enables parallel reads at the output for every pixel read at the input.
We model a line buffer as follows:
\begin{lstlisting}
type LineBuf1D = fn(Mem1D) -> Mem1D;
fn make_linebuf1d(width: int) -> LineBuf1D;
// similar for LineBuf2D
\end{lstlisting}
Akin to \lst|Regs1D| (see \cref{subsec:anyhls-core-mem}), a recursive call builds an array of line buffers
(each line buffer will be declared by a separate memory component in the residual program similar to on-chip array in \cref{fig:anyhls:memory:abstractions}).

\subsubsection{Sliding Window}\label{sec:library:swin}
Registers are the most amenable resources to hold data for highly parallelized access.
A sliding window of size $w \times h$ updates the constituting shift registers by a new column of $h$~pixels and enables parallel access to~$w \cdot h$ pixels.

\noindent\makebox[\linewidth]{%
\scalebox{0.85}{\usebox\anyhlsSlidingWindow}%
}

\noindent
This provides high data reuse for temporal locality and avoids waste of on-chip memory blocks that might be utilized for a similar data bandwidth.
Our implementation uses \lst{make_regs2d} for an explicit declaration of registers and supports pixel-based indexing at the output.
\begin{lstlisting}[
  float=!t,
  belowskip=-10pt,
  ]
type Swin2D = fn(Mem2D) -> Mem2D;
fn @ make_sliding_window(w: int, h: int) -> Swin2D {
  let win = make_regs2d(w, h);
  // ...
}
\end{lstlisting}
This will instantiate $w \cdot h$ registers in the residual program, as explained in \cref{subsec:anyhls-core-mem}.

\subsection{Loop Abstractions for Image Processing}\label{sec:library:loop-abstractions}
\secskip
\subsubsection{Point Operators}
Algorithms such as image scaling and color transformation calculate an output pixel for every input pixel.
The point operator abstraction (see \cref{lst:point-op}) in \anyhls yields a vectorized pipeline over the input and output image.
This abstraction is parametric in its vector factor~\lst|v| and the desired operator function~\lst|op|.
\begin{lstlisting}[
  label={lst:point-op},
  caption={Implementation of the point operator abstraction.}
  ]
type PointOp = fn(Mem1D) -> Mem1D;
fn @ make_point_op(v: int, op: Op) -> PointOp {
  @ |img, out| {
     for idx in pipeline(1, 0, img.size) {
       img.update(idx);
       for i in unroll(0, v) {
         out.write(i, op(img.read(i)));
       }
       out.update(idx);
     }
  }
}
\end{lstlisting}
\bigskip

\noindent
The total latency is
\begin{align}
{L} & =  L_\mathit{arith} +  \ceil{\nicefrac{W}{v}} \cdot H \text{ cycles}
\label{eq:point-latency}
\end{align}
where $W$ and $H$ are the width and height of the input image, and $L_\mathit{arith}$ is the latency of the data path.

\subsubsection{Local Operators}~\label{sec:library:local}
Algorithms such as Gaussian blur and Sobel edge detection calculate an output pixel by considering the corresponding input pixel and a certain neighborhood of it in a local window.
Thus, a local operator with a $w \times h$~window requires $w \cdot h$ pixel reads for every output.
The same $(w-1) \cdot h$ pixels are used to calculate results at the image coordinates ($x$, $y$) and ($x+1$, $y$).
This spatial locality is transformed into temporal locality when input images are read in raster order for burst mode, and subsequent pixels are sequentially processed with a streaming pipeline implementation.
The local operator implementation in \anyhls (shown in \cref{lst:local-op}) consists of line buffers and a sliding window to hold dependency pixels in on-chip memory and calculates a new result for every new pixel read.
\medskip

\noindent\makebox[\linewidth]{%
\scalebox{0.61}{\usebox\anyhlsLocalOpDetail}%
}

\noindent
This provides a throughput of $v$ pixels per clock cycle at the cost of an initial latency ($v$ is the vectorization factor)
\begin{equation}
  L_\mathit{initial} = L_\mathit{arith}
  + (\floor{\nicefrac{h}{2}} \cdot \ceil{\nicefrac{W}{v}} + \floor{\nicefrac{\ceil{\nicefrac{w}{v}}}{2}})
  \label{eq:latency-local-arith}
\end{equation}
that is spent for caching neighboring pixels of the first calculation.
The final latency is thus:
\begin{equation}
  {L} = L_\mathit{initial} + (\ceil{\nicefrac{W}{v}} \cdot H)
  \label{eq:latency-local}
\end{equation}

\begin{lstlisting}[
	float={!t},
  label={lst:local-op},
  caption={Implementation of the local operator abstraction. }
  ]
type LocalOp = fn(Mem1D) -> Mem1D;
fn @ make_local_op(v: int, op: Op, ext: Extents,
                   bh_lower: FnBorder,
                   bh_upper: FnBorder) -> LocalOp {
  @ |img, out| {
    let mut (col, row, idx) = (0, 0, 0);
    let wait = /* initial latency */
    let fsm = make_fsm();
    fsm.add(Read, || img.update(idx), || Compute);
    fsm.add(Compute, || {
      line_buffer.update(col);
      sliding_window.update(row);
      col_sel.update(col);
      for i in unroll(0, v) {
        out.write(i, op(col_sel.read(i)));
      }
    }, || if idx > wait { Write } else { Index });
    fsm.add(Write, || out.update(idx-wait-1), || Index);
    fsm.add(Index, || {
      idx++; col++;
      if col == img_width { col=0; row++; }
    }, || if idx < img.size { Read } else { Exit });
    fsm.run_pipelined(Read, 1, 0, img.size);
  }
}
\end{lstlisting}
Compared to the local operator in \cref{fig:teaser}, we also support boundary handling.
We specify the extent of the local operator (filter size / 2) as well as functions specifying the boundary handling for the lower and upper bounds.
Then, row and column selection functions apply border handling correspondingly in $x$- and $y-$directions by using one-dimensional multiplexer arrays similar to~\textcite{ASAP17}.

\section{Evaluation and Results}\label{sec:results}
\secskip
In the following, we compare the \ac{PPnR} results using \anyhls and other state-of-the-art domain-specific approaches including Halide-HLS~\cite{pu2017programming} and \hipacc~\cite{hipaccIccad}.
The generated HLS codes are compiled using Intel FPGA SDK for OpenCL 18.1 and Xilinx Vivado HLS 2017.2 targeting a Cyclone~V GT 5CGTD9 \ac{FPGA} and a Zynq XC7Z020 \ac{FPGA}, repectively.

The generated hardware designs are evaluated for their throughput, latency, and resource utilization.
\acp{FPGA} possess two types of resources:
\begin{enumerate*}[label=(\roman*)]
    \item computational: \acp{LUT} and \ac{DSP} blocks;
    \item memory: \acp{FF} and on-chip memory (\ac{BRAM}/M20K).
\end{enumerate*}
A SLICE/ALM is comprised of look-up tables (LUTs) and flip flops, thus indicate the resource usage when considered with the \ac{DSP} block and on-chip memory blocks.

The implementation results presented for Vivado \ac{HLS} feature only the kernel logic, while those by Intel OpenCL include PCIe interfaces.
The execution time of an FPGA circuit (Vivado \ac{HLS} implementation) equals to $T_\mathit{clk}$ $\cdot$ latency, where $T_\mathit{clk}$ is the clock period of the maximum achievable clock frequency (lower is better).
We measured the timing results for Intel OpenCL by executing the applications on a Cyclone V GT 5CGTD9 FPGA.
This is the case for all analyzed applications.
We have no intention nor license rights~\cite[\S 4]{xilinx_lic}~\cite[\S 2]{altera_lic} to benchmark and compare the considered \ac{FPGA} technologies or \ac{HLS} tools.

\secskip
\subsection{Applications}
\secskip
In our experimental evaluation, we consider the following applications:
\begin{itemize}
\item \textbf{\emph{Gaussian~(Gauss)}} blurring an image with a $5\times5$ integer kernel
\item \textbf{\emph{Harris corner detector~(Harris)}} consisting of 9 kernels that resort to integer arithmetic and horizontal/vertical derivatives
\item \textbf{\emph{Jacobi}} smoothing an image with a $3\times3$ integer kernel
\item \textbf{\emph{filter chain~(FChain)}} consisting of 3~convolution kernels as a pre-processing algorithm
\item \textbf{\emph{bilateral filter (Bilateral)}}, a $5\times5$ floating-point kernel as an edge-preserving and noise-reducing function based on exponential functions
\item \textbf{\emph{mean filter~(MF)}}, a $5\times5$ filter that determines the average within a local window via 8-bit arithmetic
\item \textbf{\emph{SobelLuma}}, an edge detection algorithm provided as a design example by Intel. The algorithm consists of RGB to Luma color conversion, Sobel filters, and thresholding
  \akif{this is not within, example comes from their website}
\end{itemize}

\subsection{Library Optimizations}
\secskip
\anyhls exploits stream processing and performs implicit parallelization.
The following subsections show the impact of those optimizations.

\subsubsection{Stream Processing}\label{sec:res-stream}
\begin{figure}
    \centering
    \scalebox{0.9}{\tikzpicturedependsonfile{plots/pipelining_simple.tikz.tex}

\begin{tikzpicture}[trim axis left, trim axis right]
  \pgfplotsset{
      width=0.9\linewidth,
      height=.45\linewidth,%
      enlarge y limits=0.25,%
      symbolic y coords={10,20,25,30,40},%
      yticklabels from table={plots/pipelining.dat}{App},
      y tick label style={font=\small},
      yticklabel style={align=center},%
      ytick={10,20,30,40},%
      ytick pos=left,%
      xmin=0,%
      xtick={0,16,35,107},
      x tick label style={font=\small},
      x tick style={transparent},%
      xlabel={Execution time [ms]},%
      extra y ticks={25},
      extra y tick labels={},
      extra y tick style={
          grid=major,
          major tick length=0pt,
      },
  }
  \begin{axis}[%
      xbar=0.25pt,%
      bar width=7pt,%
      xmajorgrids,%
      legend pos=south east,%
      legend style={draw=black!50},%
      legend cell align={left},%
      ]%
      \addlegendentry{na\"ive}
      \addplot[draw=cyan!60!black,         fill=blue!80!black, bar shift=0pt] table [x expr=\thisrowno{1}, y=app, skip coords between index={0}{2}] {plots/pipelining.dat};
      \addlegendentry{streaming pipeline}
      \addplot[draw=YellowOrange!80!black, fill=red!80!black,  bar shift=0pt] table [x expr=\thisrowno{1}, y=app, skip coords between index={2}{4}] {plots/pipelining.dat};
  \end{axis}
\end{tikzpicture}}
    \caption{
        Execution time for na\"ive and streaming pipeline implementations of the Harris and FChain for an Intel Cyclone~V for images of $1024\times1024$.
    }
    \label{fig:aocl-streaming}
\end{figure}
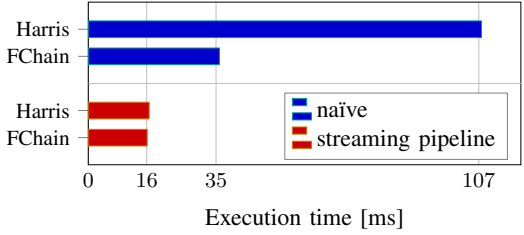

Memory transfers between \ac{FPGA}'s programmable logic and external memory are one of the most time-consuming parts of many image processing applications.
\anyhls streaming pipeline optimization passes dependency pixels directly from the producer to the consumer kernel, as explained in \cref{sec:library:stream}.
This allows pipelined kernel execution and makes intermediate images between kernels superfluous.
The more intermediate images are eliminated, the better the performance of the resulting designs.
For example, this eliminates 8~intermediate images in Harris corner and 2~in filter chain, see \cref{fig:aocl-streaming} for the performance impact.

The throughput of both streaming pipeline implementations is indeed determined by their slowest individual kernel, which is a local operator.
Consider \cref{tab:anyhls-streaming-hls}, which displays the Vivado HLS reports.
The latency results correspond to \cref{eq:latency-local}.
\begin{table}[!ht]
    \caption{
      Streaming pipeline implementations of Harris and FChain on a Xilinx Zynq.
      Data is transferred to the FPGA only once, thus similar throughputs are achieved.
      Images sizes are $1024\times1024$, $v = 1$, $f_\mathit{target}$ = 200 MHz.
    }
    \resizebox{\columnwidth}{!}{
        \begin{tabular}{lccrr}
            \toprule
            App.        & Largest mask & {Sequential Dependency} & Latency [cyc.] & Throughput [MB/s]\\
            \midrule
            FChain      & $5\times5$    & local + local + local  & 1050649        & 821 \\
            Harris      & $3\times3$    & local + local + point  & 1049634        & 825 \\
            \bottomrule
        \end{tabular}
    }
    \label{tab:anyhls-streaming-hls}
\end{table}

\secbskip
\subsubsection{Vectorization}
\begin{figure}[!h]
    \def\datafile{plots/vectorization.dat}
    \begin{center}
        \scalebox{0.9}{\tikzpicturedependsonfile{plots/vectorization_throughput.tikz.tex}

\begin{tikzpicture}
  \pgfplotsset{
      width=\linewidth,
      height=.48\linewidth,%
      enlarge y limits=0.1,%
      legend style={at={(0,1.2)},draw=none,fill=none,anchor=north west,legend columns=6,font=\small},
      legend cell align=left,
      yticklabel style={align=center},%
      ytick pos=left,
      xmin=0,%
      xmax=1500,%
  }
  \begin{axis}[%
      axis y line*=left,
      axis x line*=bottom,
      xmajorgrids,
      xmin=64,%
      xlabel={Throughput [MB/s]},
      y label style={font=\normalsize, yshift=-9pt},%
      ylabel={Vectorization factor ($v$)},
      symbolic y coords={1,2,4,8,16,32},
      ytick=data
      ]%
    \addplot[xbar=0.55pt,bar width=7pt,draw=YellowOrange!80!black, fill=Red!80!black] table [x expr=(1024 * 1024 * \thisrowno{2} / 1000 / \thisrowno{9}), y=coarsening] {\datafile};
    \addplot[black, dashed, color=red, very thick] coordinates {(1344.80,1) (1344.80,32)};
    \legend{Memory Bound [MB/s]}
  \end{axis}
\end{tikzpicture}}
        \scalebox{0.9}{\tikzpicturedependsonfile{plots/vectorization_area.tikz.tex}

\definecolor{hotorange}{cmyk}{0.000,0.510,0.930,0.100}
\definecolor{nvidiagreen}{cmyk}{0.366,0.000,0.852,0.282}
\definecolor{intelblue}{cmyk}{0.9524,0.4286,0.0000,0.3412}
\definecolor{lightgray}{cmyk}{0.1000,0.0700,0.0500,0.0000}
\definecolor{codesignred}{cmyk}{0.000,1.000,1.000,0.200}

\begin{tikzpicture}[trim axis left,trim axis right]
  \pgfmathsetmacro{\plotwidth}{\linewidth}
  \pgfmathsetmacro{\xtickcount}{6}
  \pgfmathsetmacro{\barwidth}{(\plotwidth * 0.39) / (\xtickcount * 2)}

  \pgfplotsset{
    width=\linewidth,
    height=.55\linewidth,
    enlarge x limits=0.12,
    legend style={draw=none, fill=none,font=\small},
    symbolic x coords={1,2,3,4,5,6},
    clip mode=individual,
    xticklabel style={align=center},%
    xtick=data,%
    xticklabels from table={\datafile}{coarsening},
  }

  \begin{axis}[
      axis y line*=left,
      axis x line*=bottom,
      ybar=2pt,
      bar width=\barwidth pt,
      legend style={legend columns=0, nodes={inner sep=3pt}},
      legend pos=north east,
      ylabel near ticks,
      ylabel={Resource Usage in \%},
      xlabel={Vectorization factor ($v$)},
      ymin=15,
      ymax=35,
      major x tick style = transparent,
      y label style={font=\normalsize},
      x label style={font=\normalsize},
    ]
    \addplot+[ybar, color=intelblue!60] table [x=xtick, y expr=(100 * (\thisrowno{3} / 1220  ))] {\datafile}; %
    \addplot+[ybar, color=hotorange!80] table [x=xtick, y expr=(100 * (\thisrowno{4} / 113560))] {\datafile}; %
    \legend{On-Chip Mem Blocks, Logic Resources}
    \pgfplotsinvokeforeach{1,2,3,4,5,6}{\coordinate(l#1)at(axis cs:#1,0);}
  \end{axis}
\end{tikzpicture}}
    \end{center}
    \caption{
      \ac{PPnR} results of \anyhls's mean filter implementation on an Intel Cyclone~V.
      The memory bound of the device for our setup is 1344.80 MB/s.
    }
  \label{fig:vectorization}
\end{figure}
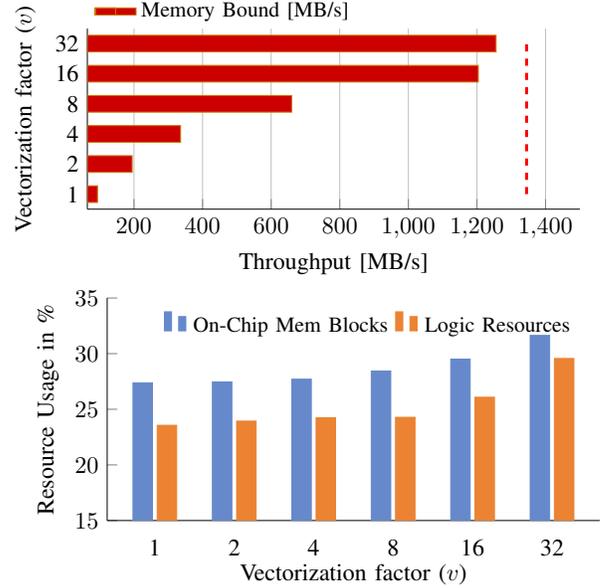

Many FPGA implementations benefit from parallel processing in order to increase memory bandwidth.
\anyhls implicitly parallelizes a given image pipeline by a \emph{vectorization factor} $v$. %
As an example, \cref{fig:vectorization} shows the \ac{PPnR} results, along with the achieved memory throughput for different vectorization factors for the mean filter on a Cyclone~V.
The memory-bound of the Cyclone~V is reported by Intel's diagnosis tool.
The speedup is almost linear, whereas resource utilization is sub-linear to the vectorization factor, as \cref{fig:vectorization} depicts.
AnyHLS exploits the data reuse between consecutive iterations of the local operators.
Data is read and written with the vectorized data types.
The line buffers and the sliding window are extended to hold dependency pixels for vectorized processing.
Thus, only the datapath is replicated instead of the whole accelerator implementation (see~\cref{sec:library:vectorization}).
All the considered applications except Bilateral in \cref{fig:aocl-results-throughput} reach the memory bound.
Bilateral is compute-bound due to its large number of floating-point operations.

\subsection{Hardware Design Evaluation}
\secskip
We evaluate the generated hardware designs based on their throughput, latency, and resource utilization.
As a reference, we use the designs generated by Halide-HLS~\cite{pu2017programming} and \hipacc~\cite{hipaccIccad}, two state-of-the-art image processing \acp{DSL} that generate better results than previous approaches (\eg Xilinx OpenCV).
In contrast to these, which implement dedicated HLS code generators, AnyHLS is essentially implemented as a library within the AnyDSL framework, as illustrated in \cref{fig:dsl_flows}.
Our focus is to show that higher-order abstractions, together with partial evaluation, are powerful enough to design a library targeting different HLS compilers.

\subsubsection{Experiments using Xilinx Vivado HLS}
We evaluate the results of circuits generated using AnyHLS in comparison with the domain-specific language approaches \hipacc and Halide-HLS.
We consider two representative applications from the Halide-HLS repository with different configurations (border handling mode and vectorization factor): Gauss and Harris.
These \acp{DSL} have been developed by FPGA experts and perform better than many other existing libraries.
The applications are rewritten for \hipacc and \anyhls by respecting their original descriptions.
This ensures that Halide-HLS applications have been implemented with adequate scheduling primitives.
\hipacc and \anyhls implementations require only the algorithm descriptions as input.

For almost all applications in \cref{tab:dsl-comp-hls,,tab:dsl-comp-hls-clamp}, \anyhls implementations demand fewer resources and deliver higher performance.
Of course, this improvement mainly stems from our library implementation.
\anyhls achieves a lower latency mainly because of the following reasons:
\begin{enumerate}[label=\roman*), leftmargin=*]
  \item
The latency of a local operator generated from \anyhls' image processing library corresponds to the theoretical latency given in \cref{eq:latency-local}, which is ${L} = L_\mathit{arith} + 1.042.442$ clock cycles for Gauss when $v=1$.
$L_{arith} = 14$ for AnyHLS' Gauss implementation as shown in \cref{tab:dsl-comp-hls}.
  \item
Halide-HLS pads input images according to the selected border handling mode (even when no border handling is defined).
This increases the input image size from ($W$, $H$) to ($W + w - 1$, $H + h - 1$), thus the latency. %
\item
\hipacc does not pad input images, but run ($H + \floor{h/2} \cdot (W + \floor{w/2})$) loop iterations for a $(W \times H)$ image and $(w \times h)$ window.
This is similar to the convolution example in the Vivado Design Suite User Guide~\cite{ug902}, but not optimal.
\end{enumerate}
\noindent
The execution time of an implementation equals to $T_\mathit{clk} \cdot \mathit{latency}$, where $T_\mathit{clk}$ is the clock period of the maximum achievable clock frequency (lower is better).
Overall, \anyhls processes a given image faster than the other \ac{DSL} implementations.

Halide-HLS uses more on-chip memory for line buffers (see \cref{sec:library:local}) compared to \hipacc and \anyhls because of its image padding for border handling.
Let us consider the number of BRAMs utilized for the Gaussian blur:
The line buffers need to hold 4 image lines for the $5\times5$ kernel.
The image width is $1024$ and the pixel size is $32$ bits.
Therefore, \anyhls and \hipacc use eight $18$K BRAMs as shown in \cref{tab:dsl-comp-hls}.
However, Halide-HLS stores $1028$ integer pixels, which require 16 $18$K BRAMs to buffer four image lines.
This doubles the number of \acp{BRAM} usage (see \cref{tab:dsl-comp-hls-clamp}).

\begin{table}
    \caption{
        \ac{PPnR} results for the Xilinx Zynq board for images of size $1020\times1020$ and $T_\mathit{target}$ = 5 ns (corresponds to $f_\mathit{target}$ = 200 MHz).
        Border handling is undefined.
    }
    \resizebox{\columnwidth}{!}{
        \begin{tabular}{lllrrrrr}
\toprule
App                     & v                  &              & \#BRAM &\#SLICE & \#DSP & Latency [cyc.]  & Throughput [MB/s] \\
\midrule
\multirow{6}{*}{Gauss}  & \multirow{3}{*}{1} & \anyhls      &      8 &    463 & 16   & 1042456 &    828.2 \\
                        &                    & Halide-HLS   &      8 &   1823 & 50   & 1052673 &    438.2 \\
                        &                    & Hipacc       &      8 &    473 & 16   & 1044500 &    764.7 \\
\cmidrule{3-8}
                        & \multirow{3}{*}{4} & \anyhls      &     16 &   1441 & 80  &  260626 &   3041.4 \\
                        &                    & Halide-HLS   &     16 &   4112 &180  &  266241 &   1640.1 \\
                        &                    & Hipacc       &     16 &   1519 & 64  &  261649 &   3064.6 \\
\midrule
\multirow{6}{*}{Harris} & \multirow{3}{*}{1} & \anyhls      &     20 &   1405 & 22  & 1041450 &    829.0 \\
                        &                    & Halide-HLS   &     16 &   2688 & 35  & 1052673 &    464.0 \\
                        &                    & Hipacc       &     20 &   1457 & 34  & 1042466 &    828.2 \\
\cmidrule{3-8}
                        & \multirow{3}{*}{2} & \anyhls      &     20 &   2513 & 44  &  520740 &  1450.4 \\
                        &                    & Halide-HLS   &     16 &   4011 & 70  &  528385 &   895.0 \\
                        &                    & Hipacc       &     20 &   2326 & 68  &  521756 &  1637.8 \\
\bottomrule
        \end{tabular}
    }
    \label{tab:dsl-comp-hls}
\end{table}
\begin{table}
    \caption{\ac{PPnR} results for the Gaussian blur with clamping at the borders.
        Image sizes are $1024\times1024$, $v = 1$, $f_\mathit{target}$ = 200 MHz.}
    \resizebox{\columnwidth}{!}{
        \begin{tabular}{lrrrrr}
\toprule
Framework   & \#BRAM & \#SLICE& \#DSP & Latency [cyc.]  & Throughput [MB/s] \\
\midrule
\anyhls     &    8  & 1646    & 16    & 1050641 & 801.8 \\
Halide-HLS  &   16  & 2096    & 50    & 1060897 & 458.7 \\
Hipacc      &    8  & 1709    & 16    & 1052693 & 820.1 \\
\bottomrule
        \end{tabular}
    }
    \label{tab:dsl-comp-hls-clamp}
\end{table}

AnyHLS use the vectorization architecture proposed in~\cite{ASAP17}.
This improves the use of the registers compared to Hipacc and Halide.

The performance metrics and resource usage reported by Vivado HLS correlate with our Impala descriptions, hence we claim that the HLS code generated from AnyHLS' image processing library does not entail severe side effects for the synthesis of Vivado HLS.
\hipacc and Halide-HLS have dedicated compiler backends for HLS code generation.
These can be improved to achieve similar performance to AnyHLS.
However, this is not a trivial task and prone to errors.
The advantage of AnyDSL’s partial evaluation is that the user has control over code generation.
Extending AnyHLS' image processing library only requires adding new functions in Impala (see \cref{fig:dsl_flows}).
Our intention to compare AnyHLS with these DSLs is to show that we can generate equally good designs without creating an entire compiler backend.

\subsubsection{Experiments using Intel FPGA SDK for OpenCL (AOCL)}
\Cref{tab:comp-altera} presents the implementation results for an edge detection algorithm provided as a design example by Intel.
The algorithms consist of RGB to Luma color conversion, Sobel filters, and thresholding.
Intel's implementations consist of a single-work item kernel that utilizes shift registers according to the FPGA design paradigm.
These types of techniques are recommended by Intel's optimization guide~\cite{intelsdk} despite that the same OpenCL code performs drastically bad on other computing platforms.

\begin{table}[!ht]
    \centering
    \caption{
        \ac{PPnR} results of an edge detection application for the Intel Cyclone V.
        Image sizes are $1024\times1024$. None of the implementations use DSPs.
    }
    \resizebox{0.9\columnwidth}{!}{
        \begin{tabular}{rlrrrr}
\toprule
 v                  & Framework    &\#M10K & \#ALM   & \#DSP   & Throughput [MB/s] \\
\midrule
\multirow{3}{*}{1}  & Intel's Imp. &   290 &   23830 &   0     &    419.5 \\
                    & \anyhls      &   291 &   23797 &   0     &    422.5 \\
                    & Hipacc       &   318 &   25258 &   0     &    449.1 \\
\midrule
\multirow{3}{*}{16} & Intel's Imp. &   -   &    -    &   0     &     -    \\
                    & \anyhls      &  337  &   29126 &   0     &   1278.3 \\
                    & Hipacc       &  362  &   35079 &   0     &   1327.7 \\
\midrule
\multirow{3}{*}{32} & Intel's Imp. &   -   &    -    &   0     &     -    \\
                    & \anyhls      &  401  &   38069 &   0     &   1303.8 \\
                    & Hipacc       &  421  &   44059 &   0     &   1320.0 \\
\bottomrule
        \end{tabular}
    }
    \label{tab:comp-altera}
\end{table}

We described Intel's handwritten \emph{SobelLuma} example using \hipacc and \anyhls. %
Both \hipacc and \anyhls provide a higher throughput even without vectorization.
In order to reach memory-bound, we would have to rewrite Intel's hand-tuned design example to exploit further parallelism.
\anyhls uses slightly less resource, whereas \hipacc provides slightly higher throughput for all the vectorization factors.
Similar to \cref{fig:vectorization}, both frameworks yield throughputs very close to the memory bound of the Intel Cyclone~V.

\begin{figure}[!t]
    \centering
    \scalebox{0.8}{\tikzpicturedependsonfile{plots/scatter_ndrange.tikz.tex}

\begin{tikzpicture}[trim axis left,trim axis right]
  \tikzset{every pin/.style={pin distance=0.1em,font=\small}}
  \begin{axis}[
      width=0.98\linewidth,
      height=.7\linewidth,
      xlabel={Hardware resources (logic utilization [\%])},
      ylabel near ticks,
      ylabel={Throughput in [MPixel/s]},
      legend style={font=\small},
      legend cell align=left,
      xticklabel={\pgfmathparse{\tick*100}\pgfmathprintnumber{\pgfmathresult}},
      ymode=log,
      scatter/classes={%
      n={mark=triangle*,red},%
      y={mark=*,gray!50}}]

    \addplot[scatter, only marks, mark options={scale=1.3}, scatter src=explicit symbolic]
       table[x expr=(\thisrowno{3} / 113560), y expr=(1024 * 1024 / 1000 / \thisrowno{8}), meta=Altera] {plots/ndrange.dat}
        node[pos=0/25, pin=right:1]{}
        node[pos=1/25, pin=right:2]{}
        node[pos=2/25, pin=right:4]{}
        node[pos=3/25, pin=right:8]{}
        node[pos=4/25, pin=right:16]{}
        node[pos=5/25, pin=above right:CU1/SIMD1]{}
        node[pos=20/25, pin=above right:CU4/SIMD16]{}
        node[pos=24/25, pin=left:CU16/SIMD1]{};
    \legend{\anyhls,NDRange}
  \end{axis}
\end{tikzpicture}}
    \caption{
        Design space for a $5\times5$ mean filter using an NDRange kernel (using the \lst|num_compute_units| / \lst|num_simd_work_items| attributes) and \anyhls (using the vectorization factor~$v$) for an Intel Cyclone~V.
    }
    \label{fig:scatter}
\end{figure}
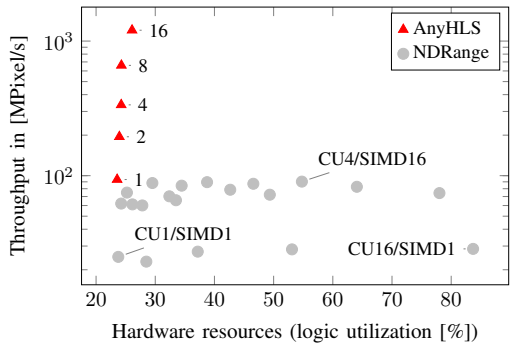

The OpenCL NDRange kernel paradigm conveys multiple concurrent threads for data-level parallelism.
OpenCL-based HLS tools exploit this paradigm to synthesize hardware.
\ac{AOCL} provides attributes for NDRange kernels to transform its iteration space.
The \lst|num_compute_units| attribute replicates the kernel logic, whereas \lst|num_simd_work_items| vectorizes the kernel implementation\footnote{These parallelization attributes are suggested in~\cite{intelsdk} for NDRange kernels, not for the single-work item kernels using shift registers such as the edge detection application shown in \cref{tab:comp-altera}.}.
Combinations of those provide a vast design space for the same NDRange kernel.
However, as \cref{fig:scatter} demonstrates, \anyhls achieves implementations that are orders of magnitude faster than using attributes in \ac{AOCL}.

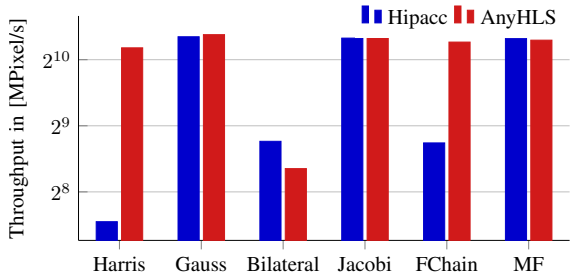
\begin{figure}[!t]
    \begin{center}
        \def\anyhlsdatafile{plots/anyhls/throughput_aocl.data}
        \def\hipaccdatafile{plots/hipacc/throughput_aocl.data}
        \scalebox{0.8}{\tikzpicturedependsonfile{plots/throughput_comp_aocl_vert.tikz.tex}

\definecolor{codesignred}{cmyk}{0.000,1.000,1.000,0.200}
\definecolor{hotorange}{cmyk}{0.000,0.510,0.930,0.100}
\definecolor{ultimatered}{cmyk}{0.040,0.890,0.620,0.030}
\definecolor{nvidiagreen}{cmyk}{0.366,0.000,0.852,0.282}
\definecolor{intelblue}{cmyk}{0.9524,0.4286,0.0000,0.3412}
\definecolor{armblue}{cmyk}{0.9006,0.1813,0.0000,0.3294}
\definecolor{darkgray}{cmyk}{0.0000,0.0000,0.0000,0.6000}
\definecolor{lightgray}{cmyk}{0.1000,0.0700,0.0500,0.0000}
\definecolor{purple}{cmyk}{0.2105,0.8565,0.0000,0.1804}
\definecolor{whitesmoke}{cmyk}{0.0000,0.0000,0.0000,0.0392}

\begin{tikzpicture}[trim axis left,trim axis right]
 \pgfplotsset{
		width=1.1\linewidth,
		height=.6\linewidth,
    enlarge y limits=0.1,
    ymajorgrids,
    xtick=data,
    xticklabels from table={\anyhlsdatafile}{App},
    compat=1.3
}
	\begin{axis}[
    ybar,
    axis lines*=left,
    bar width=10pt,
    log basis y={2},
    ymode=log,
		xmode=normal,
    legend entries={Hipacc, AnyHLS},
    legend columns=2,
    legend style={at={(1,0.9)},anchor=south east,draw=none,nodes={inner sep=3pt}},
    ylabel={Throughput in [MPixel/s]},
  ]
    \addplot+[ybar, color=blue!80!black]
      table [y expr=((1024 * 1024 * \thisrowno{2}) / 1000 / \thisrowno{4}), x=ytick]  {\hipaccdatafile};
    \addplot+[ybar, color=codesignred!90]
      table [y expr=((1024 * 1024 * \thisrowno{2}) / 1000 / \thisrowno{4}), x=ytick]  {\anyhlsdatafile};
	\end{axis}

\end{tikzpicture}}
    \end{center}
    \caption{Throughput measurements for an Intel Cyclone~V for the implementations generated from \anyhls and \hipacc. Resource utilization for the same implementations are shown in \cref{fig:aocl-results-area}.}
    \label{fig:aocl-results-throughput}
\end{figure}

Finally, \cref{fig:aocl-results-area} and \cref{fig:aocl-results-throughput} present a comparison between \anyhls and the \ac{AOCL} backend of \hipacc~\cite{fpl16}.
As shown in \cref{fig:dsl_flows}, \hipacc has an individual backend and template library written with preprocessor directives to generate high-performance OpenCL code for FPGAs.
In contrast, the application and library code in \anyhls stays the same.
The generated AOCL code consists of a loop that iterates over the input image.
Compared to \hipacc, \anyhls achieves similar performance but outperforms \hipacc for multi-kernel applications such as the Harris corner detector.
This shows that \anyhls optimizes the inter-kernel dependencies better than \hipacc (see \cref{sec:library:stream}).

\begin{table}[!t]
    \centering
    \caption{\ac{PPnR} for the Intel Cyclone~V. Missing numbers (-) indicate that the generated implementations do not fit the board.}
    \resizebox{0.9\columnwidth}{!}{
        \begin{tabular}{lrlrrrr}
\toprule
 App                    &  v    & Framework &\#M10K & \#ALM      &  \#DSP & Throughput [MB/s]\\    \\
\midrule
\multirow{2}{*}{Gauss}  &  16   & \anyhls   &  401  &   37509    &   0   &  1330.1 \\
                        &  16   & \hipacc   &  402  &   35090    &   0   &  1301.2 \\
\midrule
\multirow{2}{*}{Jacobi} &  16   & \anyhls   &  370  &    31446   &   0   &  1328.8 \\
                        &  16   & \hipacc   &  372  &    30296   &   0   &  1282.9 \\
\midrule
\multirow{2}{*}{Bilat.} &  1    & \anyhls   &  399  &    79270   &  153  &  326.6 \\
                        &  1    & \hipacc   &  422  &    79892   &  159  &  434.7 \\
\midrule
\multirow{3}{*}{MF}     &  16   & \anyhls   &  400  &    39266   &   0   &  1255.68\\
                        &  16   & \hipacc   &   -   &     -      &   -   &   -     \\
                        &   8   & \hipacc   &  351  &    31796   &   0   &  1275.9 \\
\midrule
\multirow{2}{*}{FChain} &  8    & \anyhls   &  418  &    44807   &   0   &  1230.6\\
                        &  8    & \hipacc   &  645  &    64225   &   0   &  427.4 \\
\midrule
\multirow{2}{*}{Harris} &  8    & \anyhls   &  442  &    50537   &   96  &  1158.5\\
                        &  8    & \hipacc   &  668  &    74246   &   96  &  187.14\\
\bottomrule
        \end{tabular}
    }
    \label{fig:aocl-results-area}
\end{table}

\secbskip
\section{Conclusions}\label{sec:conclusion}
\secskip

In this paper, we advocate the use of modern compiler technologies for high-level synthesis.
We combine functional abstractions with the power of partial evaluation to decouple a high-level algorithm description from its hardware design that implements the algorithm.
This process is entirely driven by code refinement, generating input code to HLS tools, such as Vivado HLS and \ac{AOCL}, from the same code base.
To specify important abstractions for hardware design, we have introduced a set of basic primitives.
Library developers can rely on these primitives to create domain-specific libraries.
As an example, we have implemented an image processing library for synthesis to both Intel and Xilinx FPGAs.
Finally, we have shown that our results are on par or even better in performance compared to state-of-the-art approaches.

\section*{Acknowledgments}
This work is supported by the Federal Ministry of Education and Research (BMBF) as part of the Metacca, MetaDL, ProThOS, and REACT projects as well as the Intel Visual Computing Institute (IVCI) at Saarland University.
It was also partially funded by the Deutsche Forschungsgemeinschaft (DFG, German Research Foundation) -- project number 146371743 -- TRR 89 ``Invasive Computing''.
Many thanks to our colleague Puya Amiri for his work on the pipeline support.

\printbibliography

\end{document}